\documentclass[12pt,preprint]{aastex}

\newcommand{\kms}{km~s$^{-1}$}

\newcommand{\ncand}{447~}

\newcommand{\nvr}{112~}

\begin{document}

\title{The Outer Halo of the Milky Way as Probed by RR Lyr Variables from the Palomar
Transient Facility 
\altaffilmark{1}}

\author{Judith G. Cohen\altaffilmark{2}, Branimir Sesar\altaffilmark{2,3}, 
   Sophianna Bahnolzer\altaffilmark{4}, Kevin He\altaffilmark{2}, 
Shrinivas R. Kulkarni\altaffilmark{2}, Thomas A. Prince\altaffilmark{2},
Eric Bellm\altaffilmark{2,5},  Russ R. Laher\altaffilmark{6} }

\altaffiltext{1}{Based in part on observations obtained at the
W.M. Keck Observatory, which is operated jointly by the California
Institute of Technology, the University of California, and the
National Aeronautics and Space Administration.}

\altaffiltext{2}{Division of Physics, Mathematics and Astronomy,
California Institute of Technology, Pasadena, Ca., 91125,
jlc@astro.caltech.edu, khe@caltech.edu, srk@astro.caltech.edu, prince@srl.caltech.edu, ebellm@astro.caltech.edu} 

\altaffiltext{3}{Max Planck Institute for Astronomy, Konigstuhl 17, D-69117,
Heidelberg, Germany, bsesar@mpia.de}

\altaffiltext{4}{535 Pontius Ave. N., Apt. 507,
Seattle, WA, 98109,
sophianna.banholzer@gmail.com}

\altaffiltext{5}{Box 351580, University of Washington, Seattle, Washington, 98195,ecbellm@uw.edu}

\altaffiltext{6}{Infrared Processing and Analysis Center, California Institute of Technology, Pasadena, Ca., 91125}

\begin{abstract}

RR Lyr stars are ideal massless tracers that can be used to study the total mass
and dark matter content of the
outer halo of the Milky Way. This is because they are easy to find in the light
curve databases of large stellar surveys and their distances can
be determined with only knowledge of the light curve.
We present here a sample of \nvr RR Lyr beyond 50 kpc in the outer halo of the
Milky Way, excluding the Sgr streams, for which we have obtained moderate resolution spectra with
Deimos on the Keck 2 Telescope.  Four of these have distances exceeding 100 kpc.
These were selected from a much larger
set of \ncand candidate RR Lyr which were datamined using machine learning techniques
applied to the light curves of variable stars in the Palomar Transient Facility 
database.  
The observed radial velocities taken at the phase of the variable
corresponding to the time of observation were converted to systemic radial
velocities in the Galactic standard of rest.
From our sample of \nvr RR Lyr we determine the radial velocity
dispersion in the outer halo of the Milky Way to be $\sim$90 km s$^{-1}$ 
at 50~kpc falling to about 65 km s$^{-1}$ near 100~kpc once a small number of major outliers
are removed.  
With reasonable estimates of the completeness of our sample of \ncand 
candidates and assuming a spherical
halo, we find that
the stellar density in the outer halo declines as $r^{-4}$.

\end{abstract}



\clearpage

\section{Introduction}

We present initial results of a study of the outer halo of the Milky Way (henceforth MW) 
using a large sample of RR Lyr type ab (denoted as RRab) variables.
RR Lyr are old low-mass pulsating stars with distinctive light curves, 
amplitudes at $R$ of $\sim$0.5 to 1~mag, and periods of $\sim$0.4 to 0.8~days, 
which are unchanged on a timescale of years and in most cases decades or longer. 
These characteristics make them fairly easy to distinguish 
in a wide field, multi-epoch optical imaging survey if the survey cadence is suitable.  
Their most desirable
characteristic is that they are standard candles.  Accurate luminosities, which have
only a small dependence on metallicity  and period (see the discussion in
\S\ref{section_distance}),
can be inferred directly from the light curves, and these
stars, with $M_V \sim +0.6$~mag, are fairly luminous and hence can be detected at large distances.

There is a long history 
reaching back more than 30 years
of efforts to calibrate the RR Lyr period-luminosity-metallicity relation, many 
of which use the Baade-Wesselink \citep{baade26,wesselink69} 
infrared surface brightness technique
to establish an accurate distance scale.
Their subsequent use as distance indicators within the MW halo, primarily
for globular clusters and distant halo stars, also has a long history, see, e.g.
\cite{longmore86}, \cite{cohen92}, and many other early efforts. 
Early calibration efforts
\citep[see, e.g.][]{longmore90} demonstrated
the advantages of using IR photometry rather than optical photometry, specifically
lower amplitude of variation meaning that fewer epochs are required to determine
a mean magnitude, hence a luminosity.  Now with the
Spitzer IRAC camera \citep{spitzer,irac} and the WISE all sky survey  \citep{wise},
highly accurate photometry on a stable space based platform
enables even more precise distances for RR Lyr, with  recent period-luminosity calibrations
for the WISE bandpasses
carried out by \cite{madore13} and by \cite{wise_rrlyr}.  
Furthermore, the HST-Fine Guidance Sensor cameras
were used by \cite{benedict11}
 to determine trigonometric parallaxes to several of the nearest field
RR Lyr, in principle providing a fundamental calibration for all these efforts.
\cite{sesar_gaia17} used the Tycho-GAIA Astrometric Solution \citep{michalik15} parallaxes
of nearby RR Lyr
from the GAIA Data Release 1 \citep{gaia_dr1}
to verify existing period--luminosity--metallicity relationships of previous studies, illustrating the potential for very high
accuracy distances for RR Lyr stars with future GAIA releases.

Our survey for RR Lyr is focused on fundamental mode pulsators, i.e. RR Lyr type $ab$.
Type $c$ RR Lyr stars, which are first overtone pulsators, 
comprise roughly
23\% of the total RR Lyr population \citep{ogle_magellanic_2016}.
They are systematically less luminous than RRab by about 0.25~mag 
and have shorter periods \citep{m4_2015}.  RR Lyr type $c$ can be distinguished from 
fundamental mode pulsators
by their smoother, more sinusoidal light curves,
but this makes
their separation from contact binary systems more challenging.

We are now in an era of large digital imaging surveys, including 
 the SDSS \citep{sdss}, 
the Palomar Transient Facility (\citealt{law09}, \citealt{rau09}) and its successors,
the Catalina Real-Time Transient Surveys (CRTS) \citep{crts},
and the Pan-STARRS survey \citep{panstars1,panstars2}, with LSST coming in the next decade.
GAIA recently had its first data release as well \citep{gaia_dr1},
 with more to follow in due course.
These surveys, with their huge databases can,
depending on their cadences and limiting magnitudes, be used to identify ever larger samples of
ever more distant RR Lyr, continuing and expanding on much earlier efforts
\citep[see, e.g.][]{wetterer96}.  Such samples enable studies of the outer halo of the Milky
Way,  as well as of streams and substructures therein.  RR Lyr are particularly  
useful for isolating halo substructures as they stand out through their variability and blue color
against the numerous foreground Galactic disk and inner halo stars; \cite{sesar12}
and \cite{sesar_orphan}
have utilized the Palomar Transient Facility (PTF) samples for this purpose.

Our survey for RR Lyr in the outer halo of the
Galaxy carried out with the PTF begins at a heliocentric distance of 50~kpc
and reaches out to distances of $\sim$110~kpc. 
Previous surveys of halo RR Lyr stars include \cite{viv01}, \cite{keller08},
\cite{mic08}, \cite{sesar11}, \cite{sesar_linear} and \cite{drake_crts}, among others.
Our survey presents
a significant improvement over anything previously published 
in sample size and in precision of distances in the 50 to 100~kpc range.
We present in this paper the radial velocity data obtained to date for these distant RR Lyr
through moderate resolution spectroscopy at the Keck Observatory as well as  
a preliminary halo density distribution derived from our full RR Lyr sample.

An overview of the PTF is given in \S\ref{section_ptf}. The following sections 
briefly review how we derived
our RR Lyr sample, then describe how we calculate distances from the light curve parameters.
We then discuss our spectroscopic follow-up campaign at the Keck Observatory
to measure radial velocities, and present $v_r$ for \nvr\ RR Lyr in the outer halo of the MW
with heliocentric distances ranging from  50 to 109~kpc, and with median distance of 73~kpc,
which we subsequently use to derive the velocity dispersion in the outer halo of the MW.
Next we  give a description of our preliminary halo density distribution 
derived from our full sample of \ncand RR Lyr candidates.  This is
followed by a comparison of our results 
with the results of other recent studies of the outer halo of the MW,
and a summary.

\section{Overview of the Palomar Transient Facility \label{section_ptf} }

The PTF \citep{law09,rau09} (P.I. S.~R. Kulkarni of Caltech)
is a wide-area, two-band ($g$ and $R$ filters),
deep ($R~{\sim}20.6$ single-epoch, ${\sim}23$~mag co-added) survey aimed at systematic
exploration of the optical transient sky. 
The PTF ran for three years, ending Dec. 31, 2012, then transitioned to the intermediate-PTF (iPTF),
with the same goals and facilities but with a slightly different consortium   
membership.
The project uses the CFH12k mosaic
camera, with a field of view of 7.26 deg$^2$ and a plate scale of $1\arcsec$
pixel$^{-1}$, mounted on the Palomar Observatory 48-inch Samuel Oschin Schmidt
Telescope \citep{rah08}. The camera consists of two rows of six $2k\times4k$
CCDs, one of which is not active.

By the end of Sep. 2014, ${\sim}12,000$ deg$^2$ of sky had been observed by the iPTF 
in the Mould-$R$ filter\footnote{The PTF Mould-R filter is similar in shape to the 
SDSS $r$-band filter, but shifted 27~\AA\ redward.}
and ${\sim}2,300$ deg$^2$ in the SDSS $g^\prime$
filter at least 30 times each.
Observations are carried out with several cadences to support various major projects,
ranging from searches for
comets and asteroids to discovery and monitoring of distant SN.
For most of a lunation, the observations are performed in a
broad-band $R$ filter. The SDSS $g^\prime$
filter is used during the darkest nights.
 Under typical seeing conditions ($1.1\arcsec$ at the P48 Schmidt) the camera achieves a full
width at half-maximum intensity  of ${\sim}2.0\arcsec$ and $5\sigma$
limiting AB magnitudes of  20.6 in median seeing.

All PTF data taken by the Palomar Observatory 48-inch telescope are
automatically routed to two pipelines: a real-time transient detection pipeline
optimized for rapid detection of interesting objects, mostly SN, and hosted by 
the Lawrence
Berkeley National Lab, and a longer-term archival pipeline optimized for
high-precision photometry and hosted by the Infrared Processing and Analysis
Center (IPAC).
The IPAC pipeline performs final image reduction, source extraction, and
photometric and astrometric calibration (\citealt{gri10, ofe12}; Laher et
al. 2014). The photometric uncertainty provided by this pipeline is
smaller than ${\sim}0.01$ mag for $R<16$ sources and increases to 0.2 mag at
$R=20.6$. The algorithm used for photometric calibration is based on 
that of \citet{hon92} as modified by \citet{ofe11} and  by \citet{lev11}.

The PTF R photometric calibration attempts, within the limits imposed by a survey
almost all of whose imaging, especially prior to 2015, was acquired with only a R filter,
to reproduce the SDSS $r'$ system.
Relative to the reference UCAC-3 astrometric catalog \citep{zac10}, the
astrometric precision of PTF coordinates is about $0.1\arcsec$ in RA and Dec.

iPTF, with partial funding from the NSF, is in the process of transitioning to the Zwicky Transient
 Facility (ZTF), to begin operation in late 2017.  The 47 deg$^2$ field of view of the 
ZTF camera will be roughly 6 times the area of the PTF camera, and larger than  
the field of each of the photographic
plates used for the Palomar Sky Survey.   
This, combined with better CCDs with faster readout times, will enable ZTF to observe 
the sky
more than 10 times faster than PTF, while still reaching the same
magnitude limit.

\section{Sample Selection \label{section_selection} }

RR Lyr stand out in a wide field imaging survey because they are blue and variable.
However, 
the PTF is primarily dedicated to searching for explosive transients.
To optimize the cadence for this purpose
almost all PTF imaging until 2014 was carried out with the $R$ filter.
Thus for stellar broad-band colors we relied on the SDSS \citep{sdss}.  In the
SDSS the imaging and thus the derived photometry 
for a variable star  was essentially simultaneous 
for each of the 5 filters. 

We developed a  probabilistic measure  of whether
or not a star is a RR Lyr variable based largely on its light curve characteristics.
As described briefly in \cite{stacking},
we have chosen to use the random forest classifier to isolate a sample of RR Lyr variables
from the PTF data.
This is a supervised machine-learning algorithm that uses a training sample and a feature 
set to build a forest of decision trees.   
Random forest algorithms are able to determine the 
importance of each feature used for classification and they are not strongly affected by outliers. 
Random forest classifiers also tend to be less affected by small changes in the training sample than other classification trees because of the random selection of a subset input features at each
node, resulting in a selection that maintains accuracy while 
reducing correlation (\citealt{bre01}).
They are easily applied to very large sets of time series data (i.e. light curves), 
and have been  used extensively
on such data sets in the past few years, see, e.g. \cite{ric11}, \cite{macho2014},
\cite{random_kepler}, and \cite{random_qso} for QSOs,  
and in high-energy physics, see, e.g. \cite{random_gamma}.
Because of the algorithm's features and its previous success in classifying variable sources, it was selected as the algorithm of choice for the RR Lyrae classifier.

As a supervised algorithm, the random forest classifier requires a training sample and a 
set of features.  After some experimentation, we settled on 10 features to characterize the light curves, including several suggested by \cite{stetson96}. 
The training sample consisted of PTF light curves of RR Lyr stars and non-RR objects
in SDSS Stripe 82, where the RR Lyr are identified in \cite{sesar_stripe82}, and the
remaining  stellar objects are non-RR stars.
\cite{nina_ps1} gives extensive details on a similar
selection applied to the Pan-STARRS PV2 (internal process V2) data.
The output of the classifier (denoted $Pr$, range 0 to 1) is a measure of 
probability that the light curve under consideration is that
of a RR Lyr.  The rank ordering of the $Pr$ values is correct, but the conversion to an actual
probability has not been quantified.  Light curve parameters are also determined,
including the period, amplitude, and epoch of maximum light which defines $\phi = 0$.

Experience gained with the random forest classifier suggested that
30 epochs, provided that they are well spaced compared to the typical
RR Lyr period, suffice to identify a RR Lyr variable,
phase the light curve, and determine its period.  
We thus require a minimum of 30 detections 
of a given star for it to be included in the RR Lyr search.
Until the fall of 2014, after this sample was originally assembled,
the RR Lyrae project had no assigned P48 time.  
Thus to assemble our sample of candidate RR Lyr stars, we datamined the
PTF archive in 2013, searching for high galactic latitude fields
that had more than 40 R images.  We ran
the random forest classifier on the time series of photometry
(ignoring images which yielded only upper limits instead of detections) 
for all stars in such fields that showed
evidence of variability and that had 30 or more detections at R.  We retained only those with
a minimum $Pr$ of 0.70 and which have reddening corrected
$g-r$ colors from the SDSS within the range
appropriate for RR Lyr, e.g. that of \cite{sesar_stripe82}.  As a final check,
the NASA/IPAC Extragalactic Database was used
to remove known QSOs. 

The area on the sky of the Sgr stream within 
 $9\arcdeg$ of the orbital
plane of this tidal stream (i.e., ${|}B_{Sgr}{|} < 9\arcdeg$, where
$B_{Sgr}$ is the latitude in the Sgr stream coordinate system
defined in Appendix A of \citealt{belokurov14})
 was excluded.  Other known streams, compiled recently  by \cite{grillmair16}, are all
closer than 50~kpc and hence not relevant here.  

Because the RR Lyr survey with PTF and its successors is largely piggy-backing on the
various SN surveys, our sample probes widely separated randomly selected high galactic latitude 
pencil beam fields,
each 7.3 deg$^2$ in size.
Substructure effects should be minimized because of our
sparse sampling over a very large area on the sky.

\begin{figure}
\epsscale{0.9}
\plotone{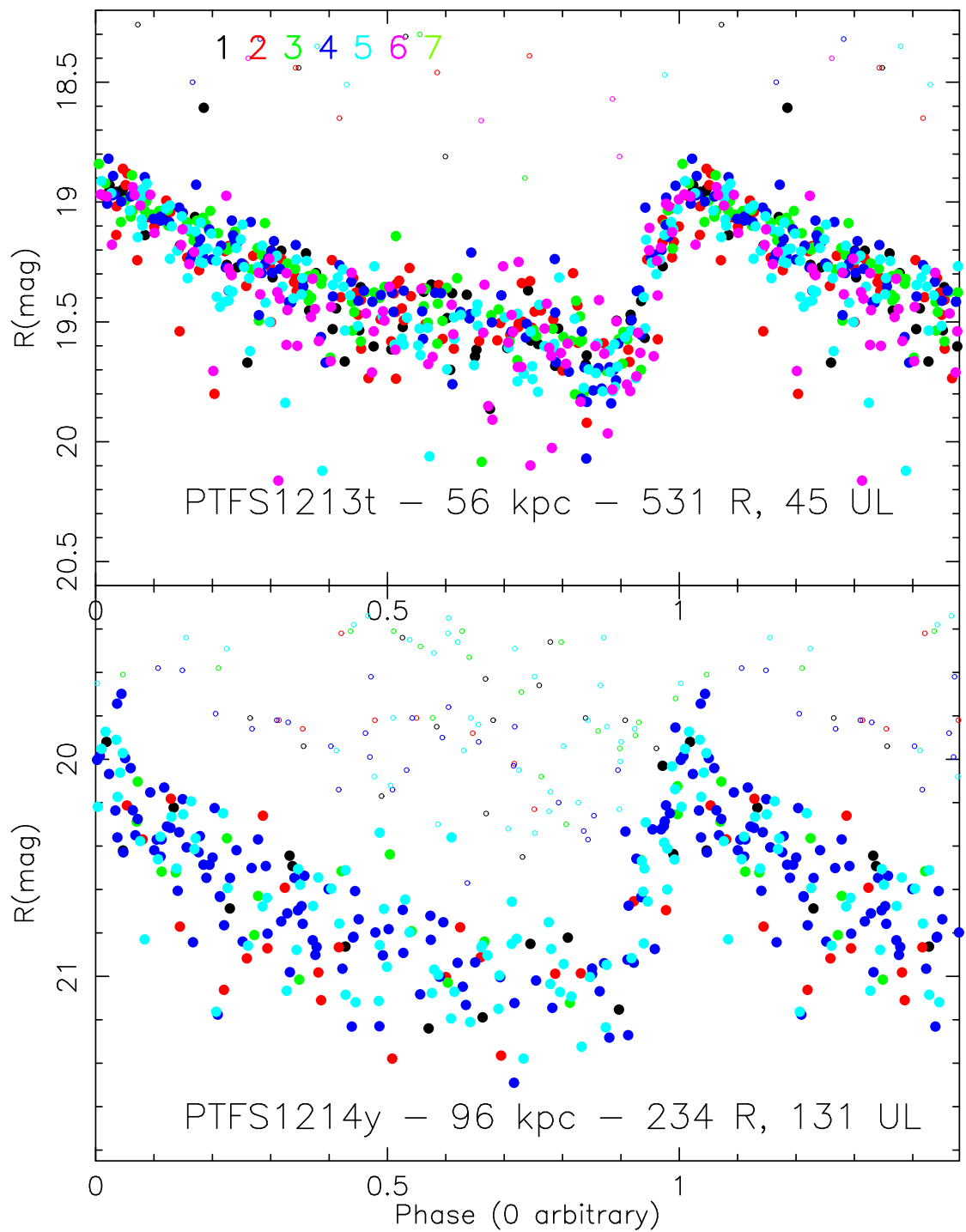}
\caption[]{Phased light curve for PTFS1213t 
(RA, Dec: 199.472902 32.118009,
$r = 56$~kpc, 576 epochs of R imaging, of
which 45 are only upper limits, indicated by small open circles) and for PTF1214y
(RA, Dec: 210.61273 39.29614, 
$r =  96$~kpc, 365 epochs of R imaging, of which 131 are only upper limits).
The colors denote the years since the first PTF observation; the key for the
colors is at the top of the upper panel, black points from the first year, etc.
\label{figure_lightcurve} }
\end{figure}

Fig.~\ref{figure_lightcurve} shows the light curve of one of the brighter RR Lyr
in our sample ($r = 56$~kpc) as well as that of one of the more distant RR Lyr ($r = 96$~kpc).
The observations extend over more than 6 years with hundreds of detections in the PTF-$R$ filter,
and with good phasing throughout.  We ignore the Blazhko effect, 
a modulation with time of the pulsation amplitude seen in some RR Lyr stars, as our light curves
in general
are not of high enough quality to detect this.

Since the PTF is primarily dedicated to the discovery and study of 
high amplitude explosive transients such
as SN, observations are carried out even when sky conditions (seeing or transparency) are not
optimal, provided it is safe to open the dome.  This means that the depth and point source image
size for an individual exposure will vary over a wide range as observations of a specific field
are accumulated
over several years.  Many of the PTF images have a limiting magnitude
much brighter than that of the median $R$ = 20.6~mag (5 $\sigma$) limit.  
Some PTF images, taken under very good conditions
(clear night, excellent seeing, excellent telescope performance)
reach deeper than the nominal limit.  Thus the quality of the light curve of a candidate variable star
is not just a function of the brightness of the star and the number of epochs available in
the PTF archive. This complicates estimating the completeness corrections in our sample.

\begin{figure}
\epsscale{0.6}
\plotone{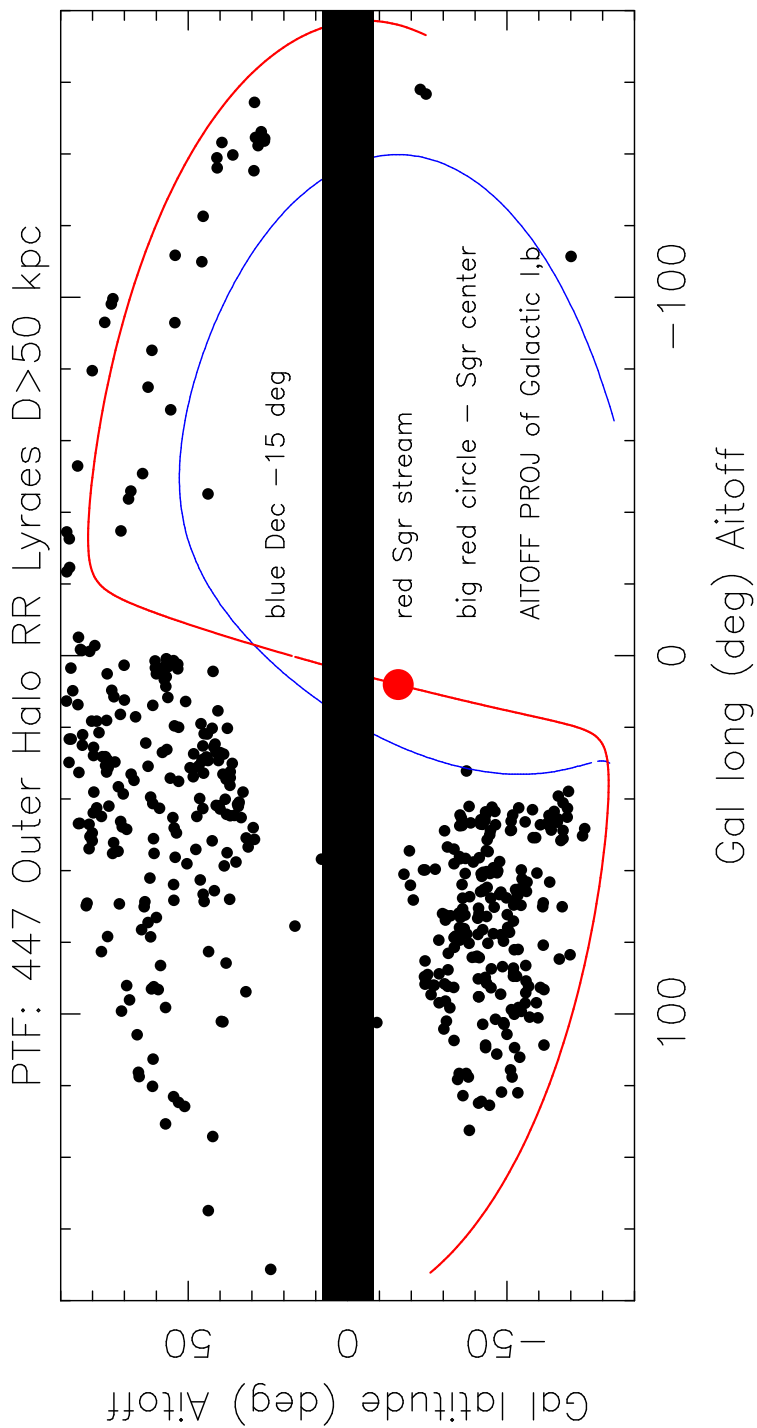}
\caption[]{The location on the sky in an Aitoff projection of galactic coordinates 
of the sample
of 447 PTF  outer halo candidate RR Lyr stars. The larger points 
have distances beyond 85~kpc.
The locus of the Sgr stream is denoted by the red curve.  The center
of the Sgr galaxy is indicated by the large red dot.
Dec $-15$~deg is indicated by the blue curve.
\label{figure_radec_all} }
\end{figure}

We assembled a list of \ncand candidate RR Lyr selected from the PTF database
to be at a heliocentric distance of 50~kpc or greater at high Galactic 
latitude\footnote{The minimum heliocentric distance of 
$r = 50$~kpc for a star in our sample
corresponds to a Galactocentric distance  
between 45.9 and 57.8~kpc depending on ($l,b$).},
with SDSS ($g - r$) colors from DR10 \citep{sdss_dr10} within an appropriate range and 
outside the Sgr streams.
Fig.~\ref{figure_radec_all} shows their location
on the sky using Galactic coordinates. 
Allowing a distance separation of 5\%, 
the closest pair of candidate RRab  has a separation on the sky of 0.16 kpc
and a distance of 50.5~kpc. There
is  one other close pair with separation on the sky of less than 0.4 kpc.

\begin{figure}
\epsscale{1.0}
\plotone{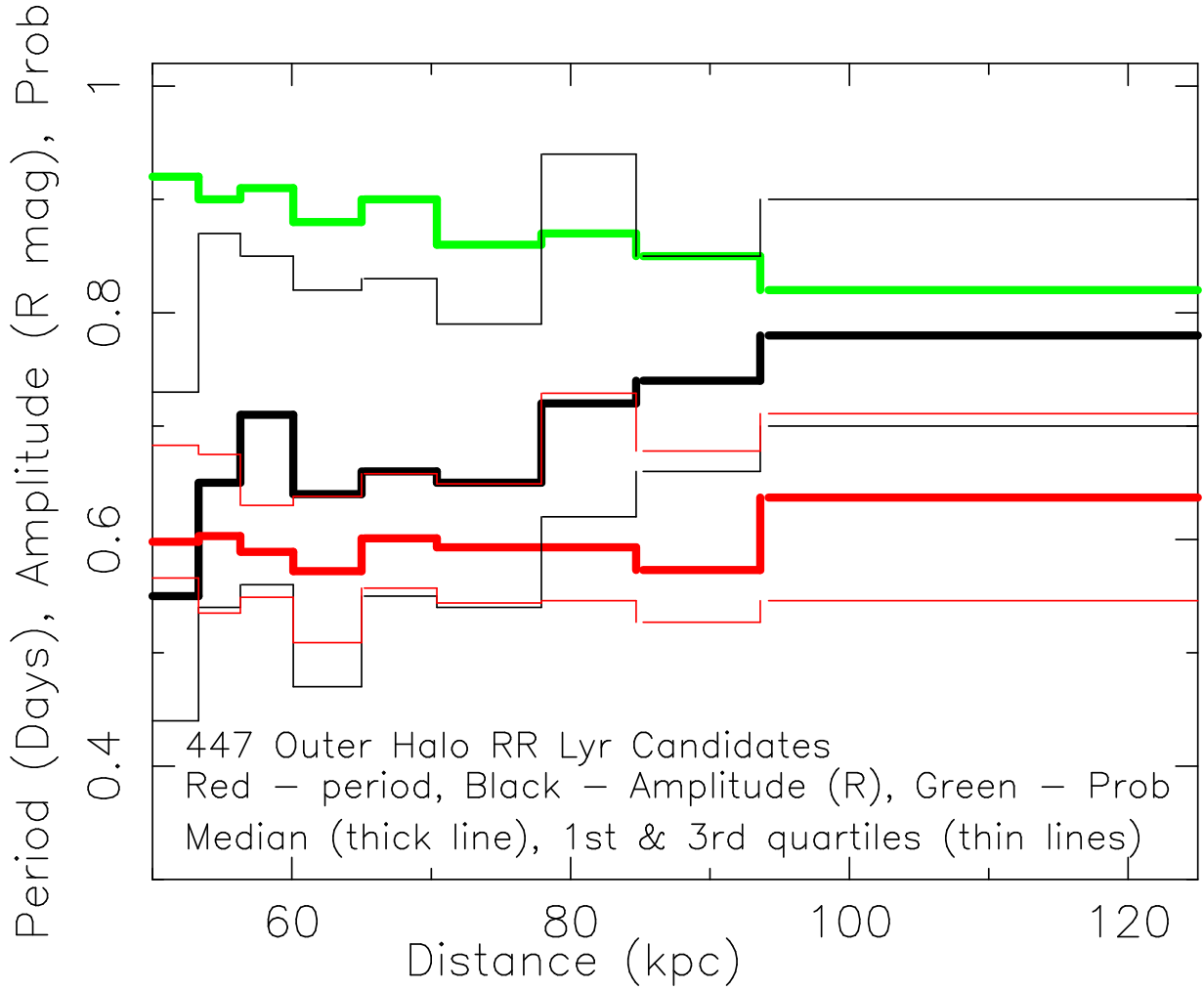}
\caption{Median, first, and third quartiles of periods and of amplitudes
of our sample of 447  RR Lyr candidates beyond 50 kpc are shown in 9 distance bins.
The median probability is also shown for each bin.  The first
bin contains 47 RR Lyr, while the more distant ones each contain 50 variables.
\label{figure_quartiles_per_amp} }
\end{figure}

Some trends appear within this sample of \ncand candidate RR Lyr
as shown in 
Fig.~\ref{figure_quartiles_per_amp}, including
a trend towards higher median amplitude with approximately constant median period
at larger distances.  There is also a trend of lower $Pr$ 
(recall that $0.7 < Pr < 1.0$) towards larger distances;
the median probability index decreases  
from 0.92 for the first bin ($r = 50$ to 53~kpc)
shown in 
Fig.~\ref{figure_quartiles_per_amp}
to 0.82 for the last bin ($r > 94$~kpc). 
These trends are not surprising given that we are approaching the limiting magnitude of the PTF survey
at the largest distances probed.

As will be described later, \nvr stars were selected 
selected for spectroscopic observations.  These consist of the higher probability
RR Lyr candidates from this list, within the constraints imposed by the specific
dates of the assigned telescope time.

\section{Distances \label{section_distance} }

\begin{figure}
\epsscale{1.0}
\plotone{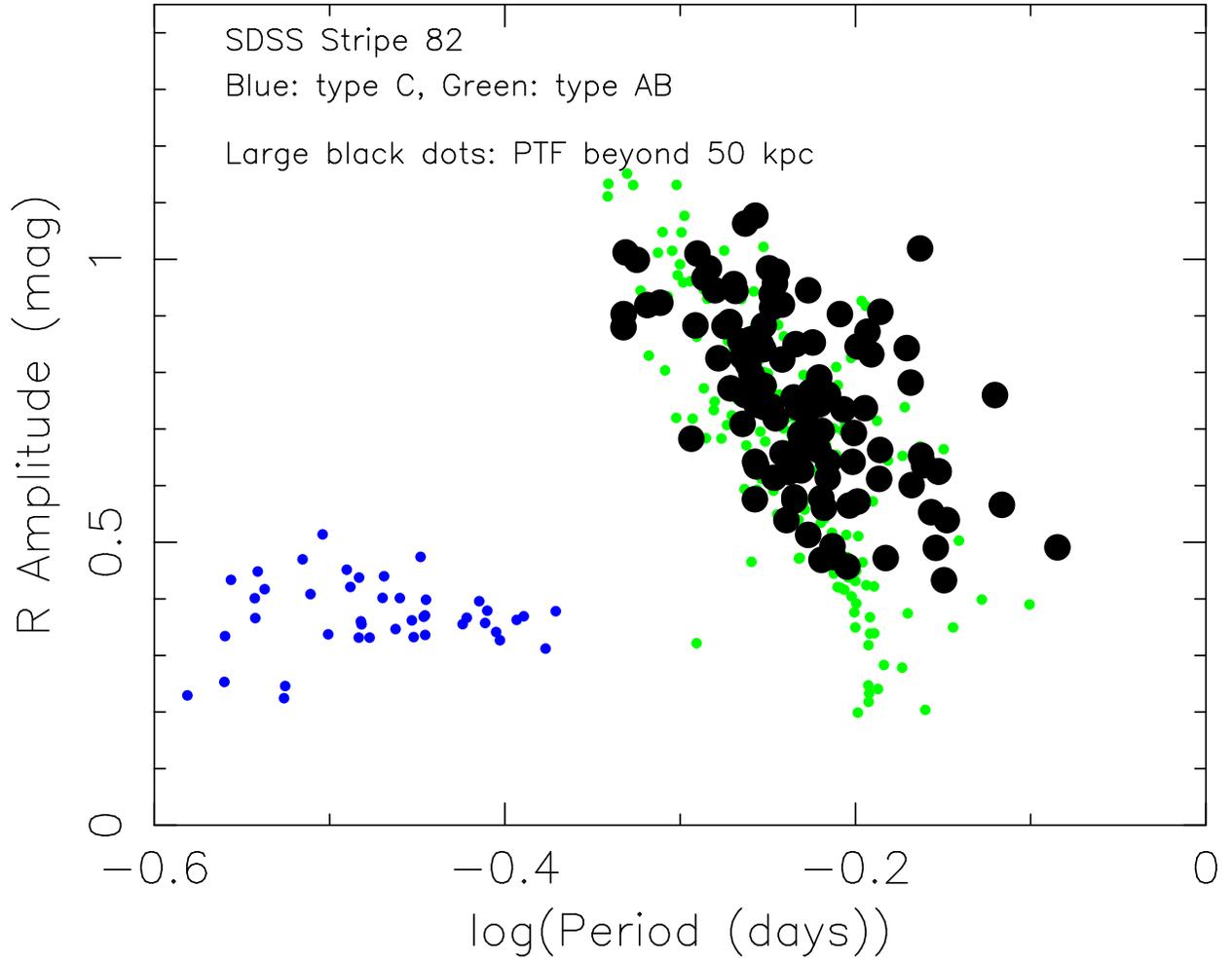}
\caption{The period -- amplitude relation for the 112 Keck PTF outer halo RR Lyr
stars is shown (large black points), as well as that of the SDSS Stripe 82
sample from \cite{sesar_stripe82} (type $ab$ in green, type $c$ in blue). 
\label{figure_period_amp} }
\end{figure}

RRab are almost standard candles, and we adopt a median for their
extinction corrected
$M_R$ mag  (averaging the  flux of the best fitting light curve template
over one period) of +0.6 mag.  However, it is well known that
there is a small dependence of luminosity on period (linear in log$P$) and on
metallicity (linear in [Fe/H]).
We first assess the range in period of RR Lyr stars.  We use the sample 
of 173 RRab isolated by \cite{sesar_stripe82} in Stripe 82 of the SDSS
with mean $R$ fainter than 17.0~mag.
This sample has excellent photometric data with many observed epochs.
It  covers a wide range in
distance and should be representative of our sample as well.
Fig.~\ref{figure_period_amp} presents the period-amplitude relation for
these stars.  The RR type c from the Stripe 82 sample 
are also shown in this figure.  Note that they have 
have shorter periods and lower amplitudes than do the RRab.

\begin{figure}
\epsscale{1.0}
\plotone{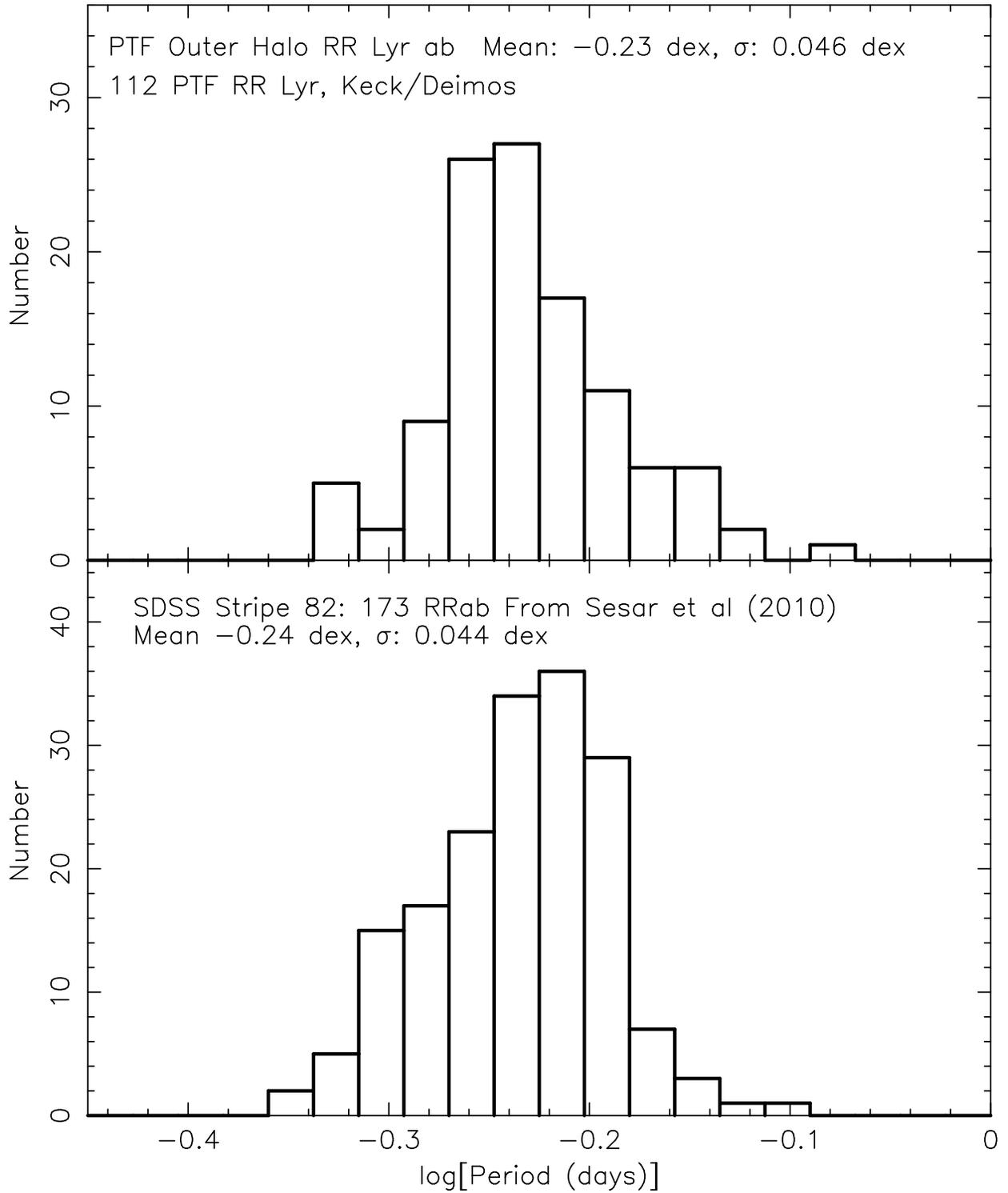}
\caption{The histogram of log(period) for our Keck $v_r$ sample is shown
in the upper panel, while that of the 173 RR Lyr~$ab$ from 
\cite{sesar_stripe82} is shown in the lower panel. 
\label{figure_period_hist} }
\end{figure}

A histogram of the periods
for the 173 RR~$ab$ stars in Stripe 82 with $R$ fainter than 17.0~mag
from the sample of \cite{sesar_stripe82}
is shown in shown in the lower panel of Fig.~\ref{figure_period_hist}; the upper panel
displays the same for our PTF RRab sample.  The values for mean and rms dispersion 
of each of the two samples are
indicated on the figure; they are essentially identical, which is gratifying, as both
probe deep into the outer halo of the Milky Way.

We correct for the period term in the luminosity of RRab adopting
the coefficient given by \cite{marconi15}, who
 present theoretical period-luminosity relations for RR Lyr
stars over a range of metallicity based on their new nonlinear time-dependent
convective hydrodynamical models of RR Lyr stars.  
These supersede earlier calculations by \cite{chaboyer99}, \cite{plz} and others.
We note note
that their dependence on log($P$) for fundamental mode RR Lyr 
in the $i$ band is 1.6 times larger than that of \cite{plz} and
therefore we may hope that use of their coefficients will provide an upper limit
to the change in mean $R$ with both $P$ and [Fe/H].

The luminosity dependence on the period, which is ${-1.39}~{log(P)}$, is in general small, as
the median period for our outer halo RR Lyr sample is 0.553 days
with ${\sigma}log(P) = 0.044$~dex.  This results in a change in
distance of less than 3\%; even at the extreme high and low values
of log($P$), 
$ P = 0.456$~days and $P = 0.793$~days,  the resulting change
in distance incurred by including the period dependent term
does not exceed 10\%.  Note that the sample of RRab
in SDSS Stripe 82 studied by \cite{sesar_stripe82} has a median
period of 0.582 days, and a $\sigma$ for log($P$) of 0.044~dex,
almost identical to that of our outer halo sample.

Corrections for interstellar absorption were applied based on the reddening
map of \cite{int_extinc}.
If a reddening map with larger
extinction at high galactic latitude is used, the distances to the RRLyr
would increase.

The absolute luminosity of RRab also depends on the metallicity, for which
[Fe/H] is used.
 \cite{halo_mdf} have established the metallicity distribution of the outer
halo for very low metallicities;  the fraction of the stellar content of the Milky Way halo which is
extremely metal poor is very small.
 Again using the coefficients of \cite{marconi15}, we find
that potential variations of [Fe/H] of $\pm$0.5 dex about a (low, but
not extremely low) mean metallicity 
 leads to an uncertainty in the distance of 4\%.

Another key issue is the accuracy of the mean $R$ mag measured from our
light curves.  Although the uncertainty of a R measurement at a single
epoch may be large, up to 0.2~mag, the mean R will be much more accurate.
A reasonable estimate of this, particularly for stars with many epochs
(ignoring upper limits)
in their light curves, is 0.03~mag, which corresponds to a distance
uncertainty of 1.5\%.

Thus if one assumes that the mean metallicity
in the outer halo  beyond 50 kpc is low and only has a modest gradient 
with distance and a modest range at any outer halo location,
which seems appropriate for the outer halo excluding the Sgr streams, 
then based on the uncertainties
found above, our distances for RRab with good light curves should be precise to 5\%.
Here the dominant term results from the unknown metallicity. 
Light curve quality for these RRab will improve with additional observations 
once ZTF is commissioned resulting
in better light curves with more  detections.

An empirical test of how large (actually how small) the distance errors might be
for RR Lyr due to their range of periods was carried out by B.~Sesar.
Using the Pan-STARRS RRLyr catalog (to become publicly available on Nov. 1, 2017)
he calculated the dispersion in distance based on assuming a fixed absolute
$R$ mag, ignoring the period and metallicity dependences, for a large
($\sim 200$) sample of RRab in the Draco dSph galaxy.  He measured a rms scatter of 0.08~mag,
corresponding to a distance precision of 4\% for this sample.  In 
\S{3.3} of \cite{sesar_ps1} this test of calculating the dispersion
in distance is extended to 
two additional dSph
satellites of the Milky Way, Sextans, and Ursa Minor, again with excellent results.
In these tests the metallicity dependence within each of these dSph galaxies
was ignored.  The metallicity range within Draco extends from [Fe/H] = $-$3.0 to $-1.5$~dex
\citep{cohen09,kirby11} and the range within Ursa Minor is similar
\citep{cohen10,kirby11}.  Thus
the metallicity dependence is not a significant contributor for most, if not all, outer halo stars.
It is clear that distances accurate to 5\% can be obtained for RRab
with high quality light curves.


The above discussion does not address the issue of the  the uncertainty 
in the adopted extinction map.  The recent extinction coefficients
of \cite{schlafly11} (their table 6) and the \cite{schlafly14} dust map
lead to somewhat higher extinction at high Galactic latitude
than that we adopt, which would result in our distances being
slightly underestimated.  Furthermore we have adopted an
absolute R mag (corrected for reddening) for the median of our sample of +0.6 mag.
If this choice is incorrect, all of our distances need to be scaled
appropriately.

%

\section{Light Curve Quality \label{section_light_curve_quality}}

Table~\ref{table_lightcurve} gives the light curve parameters for each star
in our sample of \nvr\ RRab in the Keck $v_r$ sample.
The last three columns of this table indicate for each RRab
the number of available $R$-band detections, the number of $g$-band detections,
and an assessment of the light curve quality in the PTF Stellar 
Light Curve database as of late 2016.
The latter was set 
by visual inspection by the first author in late 2016.  
Quality 1 light curves are clearly RR Lyr variables, quality 2 are probably RR Lyr,
while the nature of objects with quality 3 light curves is uncertain.
 
The initial determination of light curve parameters 
for the Keck $v_r$ sample was carried out in 2014, at which
time the number of available epochs was smaller (often considerably smaller) than at present.
Beginning in late 2014 we were allocated a total of roughly 100 hours of P48 (i.e. PTF)
time made available through the Caltech allocation to improve the light curves of the more distant stars in the
Keck $v_r$ sample.   These distant stars
have a much higher fraction of non-detections than do the brighter end of our sample, and
so need additional imaging to raise the number of detections to a level that ensures accurate characterization of the light curve parameters.  Our goal in this effort is 100 detected epochs
of R-band imaging for each of the most distant RR Lyr candidates.

In late 2016 and early 2017 the light curve
of each of the stars in the Keck $v_r$ sample was checked
to look for problems in the phasing, i.e. incorrect periods
or determinations of $\phi = 0$ caused by 
the limited data available when the light curves were first determined in 2014.
As necessary, the light curve parameters were re-determined at that time, and
the correction from the observed $v_r$ to the systemic $v_r$ described in \S\ref{section_vr}
was updated using the new ephemeris parameters.  This was a crucial
step, as the initial values were in several cases sufficiently far off that the
accumulated phase change over several years significantly affected the 
derived phase correction to the observed $v_r$. 

At the present time, as  indicated in Table~\ref{table_lightcurve}, 
only 12 stars from the Keck $v_r$ sample of \nvr stars
have less than 50 detections in $R$ or 50 detections in the $g$ filter.
Only 4 stars (4\%) of the total sample of \nvr RR Lyr are classified as
having poor light curves  (i.e. quality 3).  As described earlier,
we have until very recently had no control over the observing plan for PTF/P48 time, which is
defined by the other major projects of the PTF, especially the SN projects.  Thus
the number of epochs of observation of a given star varies from a minimum of ${\sim}30$ up
to ${\sim}700$ when a RR Lyr candidate is by chance located in a field which is of 
major interest to one of the other PTF projects.

\section{Radial Velocity Measurements \label{section_vr} }

A spectroscopic campaign to obtain radial velocities for RR Lyr candidates
began at the Keck Observatory with 
the Deimos spectrograph \citep{deimos} in the spring of 2014 following
a brief effort to use the DBSP on the Hale Telescope at the Palomar Observatory,
which yielded one useful spectrum.
RRab are pulsating periodic variable stars.  Spectroscopic observations to determine $v_r$
must be taken within the range of phase such that 
($dv_r/dt$) in the stellar
atmosphere is as small as possible.
The observing list for each night was compiled from the candidate RRab
near the meridian during the night
with the appropriate range of phase ($\phi = 0.1$ to 0.7) accessible during that time.  
Observation planning therefore required
having both a coordinate list and accurate predicted phases from the start to the
end of the night for the specific date; the phases are calculated from the periods we determined
from the PTF light curves.
Candidates with high probability index ($Pr > 0.9$) were favored, but it was sometimes necessary
to incorporate candidates with lower $Pr$ to fill
in gaps in the observing plan for a specific night.

\begin{figure}
\epsscale{1.0}
\plotone{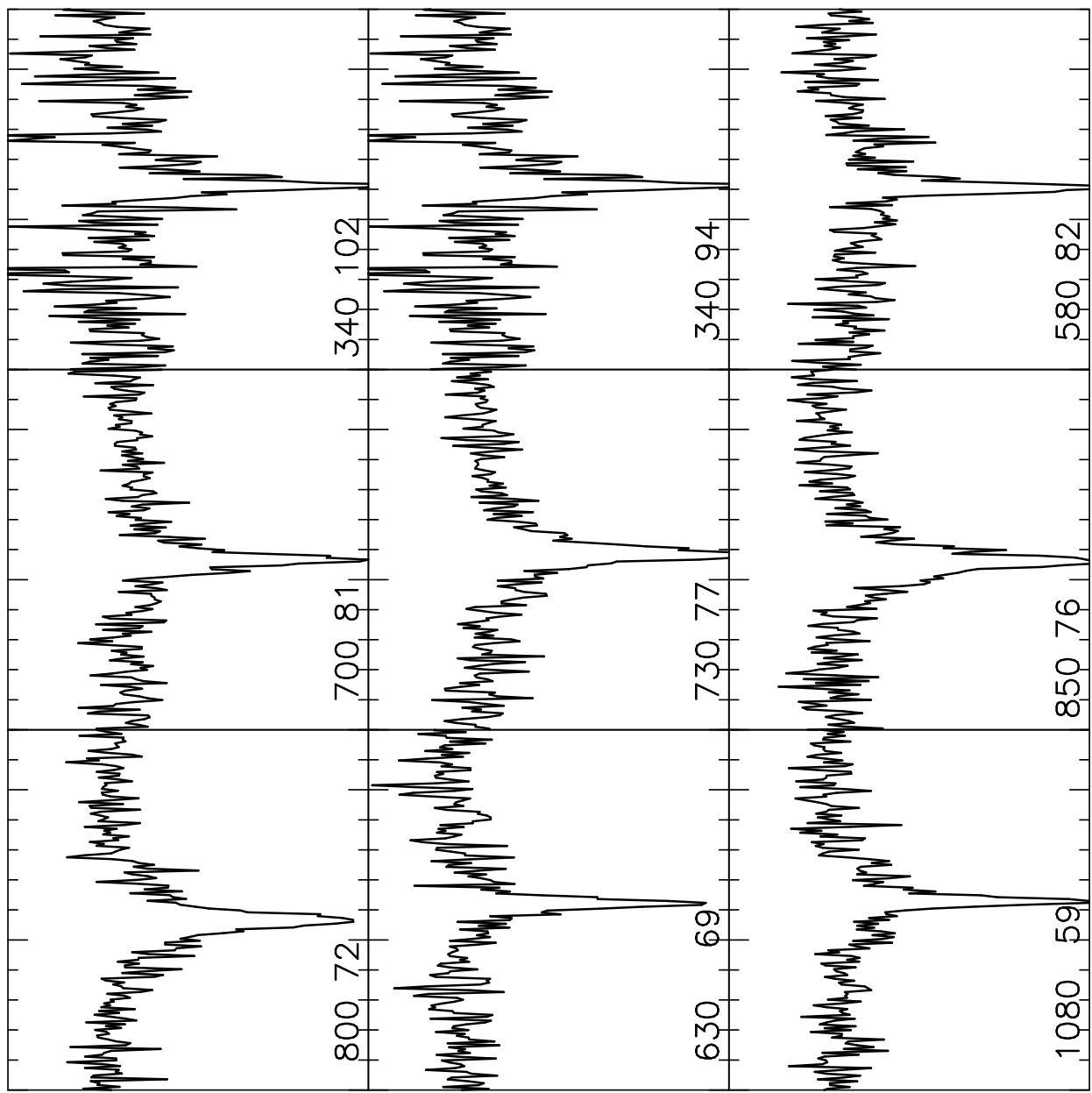}
\caption{Plots of the 1D extracted spectra for 9 RR Lyr from our sample.
The vertical scale of each panel ranges from 1.2 times the continuum
signal near H$\alpha$ to 0.5 times the continuum signal. 
The text at the bottom of each panel gives the continuum signal level
and the distance. The panels are
ordered by the distance from 59 (bottom left) to 102~kpc (upper right).
The wavelength range (X axis) of each panel is 6500 to 6620~\AA.
\label{figure_1dspectra} }
\end{figure}

The Deimos spectrograph was usually configured with
 the 600 grove/mm grating blazed at 7500~\AA\
and spectral resolution $\sim$2000 for a 1.0
arcsec wide slit and a scale of 0.65~{\AA}/pixel.  Spectra were taken with the central wavelength
set to 7500~\AA. Most spectra were taken with a 1.0 arcsec wide slit, but on nights
with good seeing, the
0.7 or the 0.8 arcsec slit was used, yielding correspondingly
higher spectral resolution.  During our first Keck run, a small number of spectra were 
acquired with the 1200 g/mm gold coated
grating blazed at 7545~\AA\  which yielded even higher spectral resolution.
However,  the
velocity precision for an exposure of a fixed time
turned out not to be better than with the 600 g/mm grating
due to the increased SNR with the 600 g/mm grating.  The maximum (and typical) exposure time
was set to 30 min to avoid excessive phase blurring. 
Fig~\ref{figure_1dspectra} shows spectra in the region of H$\alpha$ 
for 9 of our RR Lyr candidates
selected to cover the full range in distance of our sample.  Note the
degradation in the SNR at the largest distances, which arises from the fixed maximum
integration time of 30 min. 

The determination of the systemic $v_r$ for a RR Lyr requires knowledge of the phase at the time 
of the observation.  A  correction which depends on the phase is applied to get the systemic $v_r$, 
then a heliocentric correction, and finally we apply a correction to the Galactocentric rest frame (GSR).

The uncertainties in the systemic velocities include both a measurement error and a
term for the uncertainty in fitting to the model radial velocity curve.  Details for this
calculation are given in \S{5.3} of \cite{sesar12}.  Since typical $v_r$ amplitudes
over the period for
RR Lyr stars of H$\alpha$ are ${\sim}110~$km s$^{-1}$, it is important that the phase of observation
be determined accurately.  This requires accurate periods and phasing.

The primary feature we use for $v_r$ determination is H$\alpha$.
Due to the low efficiency of Deimos in the blue, we do not achieve
a SNR high enough there to use the higher
Balmer lines or the strong blue metallic lines.
\cite{sesar_vrcalib} has derived template velocity curves which
calibrate the change in velocity as a function of 
pulsation phase for several of the Balmer lines; we adopt his H$\alpha$ template here;
see also the very detailed recent study by \cite{chadid16}.
The normalized $v_r - \phi$ curves of \cite{sesar_vrcalib}
are then scaled
by the amplitude of variation of the light curve
to derive the correction from the observed $v_r$ at phase $\phi$
to the systemic velocity.
 The other major features clearly visible in these spectra are the Paschen lines around 8600~\AA,
but we are not aware of any $v_r - \phi$ calibration for them.  The infrared Ca triplet
and a few OI lines are also visible in these spectra, and we will undertake an attempt to
use them as metallicity indicators in the future.

Standard arc lamps (Ne, Ar, Kr, and Xe)
were used for  wavelength calibration, which was then tuned up slightly for each
observation of a RR Lyr candidate using
the night sky emission lines superposed on each stellar spectrum, important
as there are few arc lines in the region of H$\alpha$,
the primary feature we are using to determine $v_r$. 

From our Keck runs beginning in April 2014
and extending through Sep 2016
we have acquired spectra of roughly 135 candidate RR Lyr from our PTF sample, 
\nvr\ of which
we believe to be RRab stars based on their light curves, their colors, and their
spectra, the remainder being a few quasars, a few RR type c variables, or
other types of variable stars.  The stars in the Keck $v_r$ sample 
are widely dispersed on the sky with the closest pair in the list of candidate RRab being
separated by $0^{\circ}.2$.  Thus
no multiplexing was possible, and a substantial number of Keck nights were required
to obtain this set of spectra.  

The set of \nvr\ RR Lyr with spectroscopic $v_r$, which are listed in
Table~\ref{table_lightcurve}, have heliocentric distances of
$50 < r < 109$~kpc
with a median of $r = 73$~kpc. 
The resulting $v_r$
relative to the Galactic standard of rest and its uncertainty are given in 
Table~\ref{table_vr}.  Typical uncertainties for $v_r$ from a single
measurement range from 17 to 20~km s$^{-1}$.

  A separate list of 4 RR Lyr that were
observed during our first run with Deimos on Keck for this project, but which are probably part of
the Sgr stream, is given at the end of this table.  The exclusion region 
for the Sgr stream
was originally set to be within  5$^{\circ}$ of the orbital plane of  this tidal
stream, 
 but was raised to
9$^{\circ}$ shortly after observing commenced.
Once the exclusion region around the Sgr stream was
increased in size, these four stars were dropped from our sample of candidate 
RR Lyr in the outer halo of the Galaxy.  There are 7 stars in our sample which
are between 9 and 15$^{\circ}$ from the orbital plane of the Sgr tidal stream.

\subsection{Test of $v_r$ Accuracy}

To demonstrate the accuracy of our  systemic $v_r$ for RR Lyr variables, Table~\ref{table_vr_dup}
gives the  independent $v_r$ for those candidate RR Lyr variables from our sample
with more than one Deimos spectrum; there are 22 ($\sim$20\% of the total sample
with Keck $v_r$)
with two independent spectra.  
In several cases, the two Deimos spectra were taken
on the same night, often consecutively, but analyzed independently.  
The agreement between the two determinations of $v_r(GSR)$
for each of these 6 stars is good.  There are 11 stars
with two spectra from the same night or from consecutive nights.
These in general show small differences in $v_r$ between the two spectra.  Only three
have differences exceeding
20~km s$^{-1}$, with the largest difference being 33~km s$^{-1}$.
Given that the nominal uncertainty of a single measurement is  $\sim$20~km s$^{-1}$,
this agreement is good.

There are 8 RR Lyr candidates with two Deimos spectra taken
more than a year apart. The differences are larger here, ranging from
9 to 51~km s$^{-1}$, with two having differences exceeding 40~km s$^{-1}$.

\begin{figure}
\epsscale{1.0}
\plotone{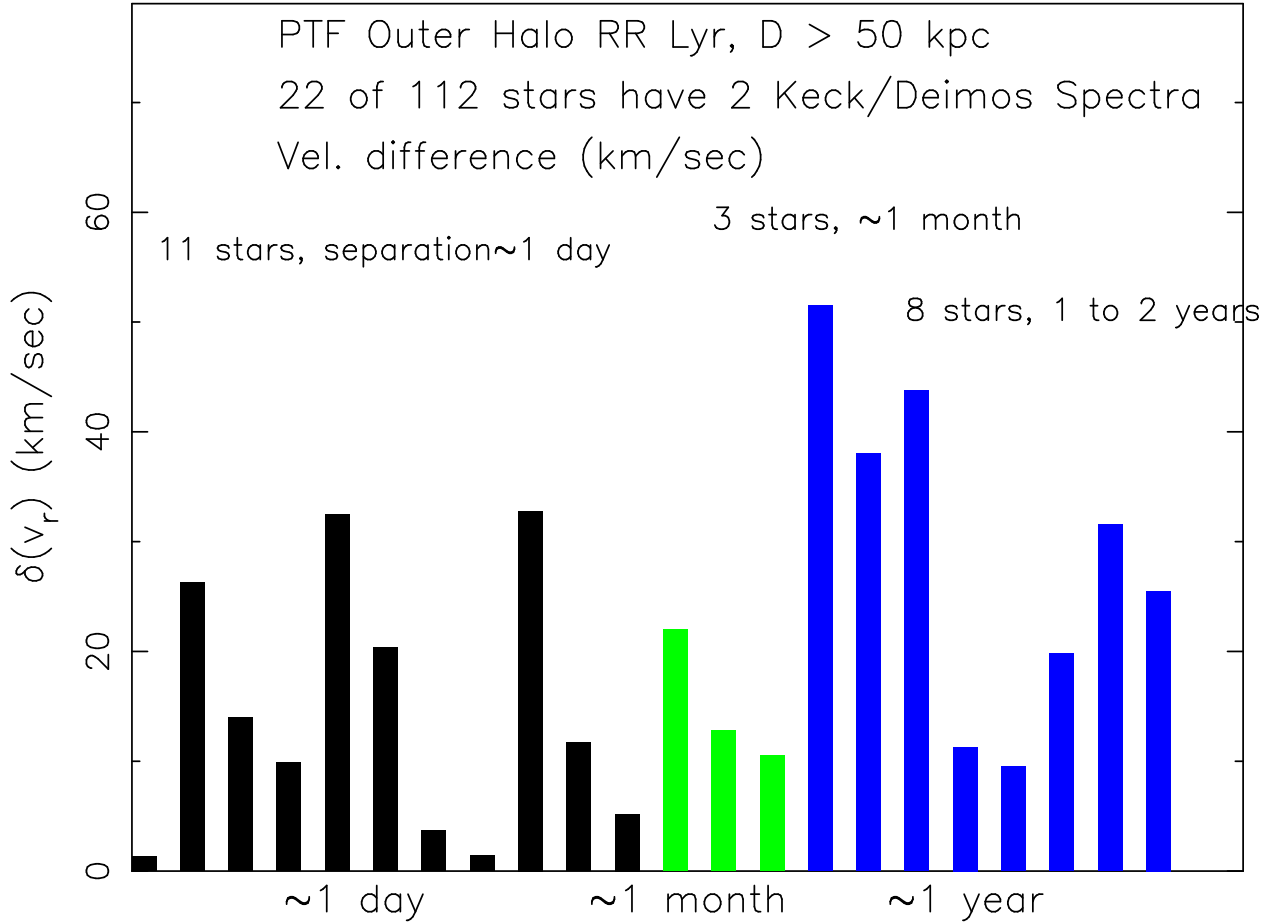}
\caption[]{The sample of 22 stars with two Keck/Deimos spectra is shown.
The vertical axis is the absolute value of the difference in $v_r(GSR)$
for each star with more than one spectrum.
The horizontal axis sorts the pairs in order of increasing separation
in time between the two observations, with difference ranging from $\sim$1 day
to $\sim$1 year.
\label{figure_vr_dup}  }
\end{figure}

The difference in $v_r$ for stars with multiple spectra is shown 
in Fig.~\ref{figure_vr_dup}
as a function 
of the separation in time between the two epochs of observation, which
increases along the X axis. As indicated above,
for small differences in time, the difference between the two $v_r$ for a given
star is within close to or within
the  expected uncertainties, but once the time interval becomes 
large (months to years),
there are two cases with disagreements exceeding 40~km s$^{-1}$ between the two derived $v_r$.

We suspect that these disagreements arise in part from possible errors in the phases due to
uncertainties in the period.
Thus our process to determine $v_r$ in the Galactic standard of rest (GSR) for these
RR Lyr variables from the observed $v_r$ using phase dependent corrections 
appears to be working reasonably well in general.  However,
for a small
fraction of our candidate RR Lyr variables, this is not the case.
While there are a few unexpectedly large discrepancies, they are relatively small compared
to the velocity dispersion among our \nvr RRLyr sample to be discussed later in
\S\ref{section_sigma}.
 
As indicated earlier, if the star were  not  a genuine RRab variable or
the key light curve parameters (period and epoch of zero phase) 
we derived were wrong, incorrect $v_r$ would be derived.
However, we note that one of the stars with two independent spectra
which shows an unexpectedly large $\delta(v_r)$
has an excellent light curve with 224 detections with the PTF-$R$ filter.

\subsection{Contaminants in the Sample of Candidate RR Lyr Stars \label{section_contaminants} }

The only blue point sources seen at high galactic latitude are RR Lyr, QSOs,
blue horizontal branch (BHB) stars and blue stragglers.  Since our primary selection
is by variability resembling that expected for  a RRab,
the non-variable  BHB stars and blue stragglers become irrelevant.
Thus the primary source of contamination is expected to be QSOs,
but the timescale and
characteristics of their variation are quite different from those of RRab. 
As the number of observed epochs of photometric monitoring
increases, and this will grow with time as the iPTF transitions into the ZTF,
the fraction of contaminating quasars will fall, because their extended 
light curves will diverge more and more from those of RRab,
the variation will not be periodic, etc.

Table~\ref{table_qso} lists the five broad lined objects (i.e. QSOs)
we have found from our spectroscopic campaign that are not included in NED.   
Given that most QSOs are 
eliminated as they do not have light
curves that resemble RR Lyr and adding in a check with NED,
 the contamination rate of QSOs within the sample selected for spectroscopy
 can be kept very low, and can, as described above,
be expected to fall with time as the survey time coverage increases.  WISE colors
\citep{wise} can also be used
to cut down the fraction of QSO contamination \citep{wise_qso} due to the difference in spectral slope between a power law and a (hot) thermal spectrum, but are of limited use for such distant objects
as they are often so faint that only the W1 color is given in the WISE catalog.

The other potential contaminant of our sample of RRab stars is 
overtone pulsators, i.e. type $c$ RR Lyr.
Several of the
variable stars listed at the end of Table~\ref{table_qso} are probably
RR Lyr~$c$.  They 
were originally believed to be RRab and were part of the spectroscopic sample, 
but as their light curves
built up with time, they became inconsistent with the period range and/or shape
appropriate for RRab and were removed. 
Given the mean difference in luminosity
at $R$ of $\sim$0.25~mag between RR Lyr pulsators in the fundamental and the first overtone modes, 
the distance of a candidate will be overestimated
by 12\% if it is actually a type $c$  rather than the more common type $AB$ RR Lyr.
Furthermore the
$v_r - \phi$ relationship for H$\alpha$ of the overtone pulsators
may be different from that derived for the RR Lyr~$ab$
that we are using.

Most importantly, type $c$ RR Lyr can be eliminated using a period-amplitude diagram, as is shown
for the SDSS Stripe 82 sample in
Fig.~\ref{figure_period_amp}.  The type $c$ variables have smaller periods
and smaller amplitudes of variation than do the fundamental mode  RR Lyr~$ab$.  
There is essentially no overlap between them in this diagram.

\section{The Radial Dependence of The Velocity Dispersion \label{section_sigma} }

Our ultimate goal is the determination of the mass of the MW out to as close
to the virial radius as possible.  We intend to use our sample of RR Lyr as massless point-source
test particles.  In support of this effort, we have ignored RR Lyr in the Sgr stream.
However, there may be previously unknown structures whose stars
may be moving with non-virialized velocities.  So before determining the velocity dispersion,
we look for evidence from our data regarding the possible presence of new substructures.
A search for previously unknown low luminosity galactic satellites 
in the outer halo that have not yet been disrupted 
was conducted by \cite{stacking} by using the RR Lyr from the PTF as 
indicators, but there was no detection, although the derived upper limit is high.  With our
$v_r$ survey we can look for evidence for the presence of more diffuse and more extended
structures.

At the large distances we probe, the dispersion of the line-of-sight velocity, which
is what we measure, is essentially identical to the dispersion of $v_r$, the radial
velocity as seen from the Galactic center.  We first consider the sample as a whole.
Fig.~\ref{figure_vr_hist} shows a histogram of the entire sample of \nvr RRab.  We see a rather
broad range spread between $-220$ and $+220$~\kms.  This may be the result of a wide
spread in $v_r$ at all Galactocentric distances,  of a trend with $r$, of the presence
of outliers due to halo structures, or of contamination in the sample with objects that are not RRab.  
As discussed in \ref{section_contaminants},
we believe that our sample has few such contaminants.

In an effort to identify whether outliers are present, we carried out an exercise
where we began with the full sample, calculating the $v_r$, ${\sigma} (v_r)$, and
the median distance.  We then removed the largest outlier in $| v_r - <v_r>(last) |$,
where ``last'' refers to  the mean $v_r$ found in the previous iteration.  We continued
doing this repeatedly. The result for $v_r$, $\sigma (v_r)$,
and the median distance for is shown in Fig.~\ref{figure_1x1} for 24 such trials.
First we note that the median distance begins for the entire sample at 74.1~kpc
and slowly decreases, ending up after 24 deletions at 71.8~kpc.  The $\sigma(v_r)$ 
for the entire sample treated as a whole begins at 98~\kms\ and ends up at 60~\kms.
It falls quite rapidly initially, suggesting the presence of some outliers, then
after about 8 RRab are removed, the decline becomes more gradual.  At the same time,
the mean $v_r$ starts at about $-19$~\kms, and rises to +3~\kms\ at the end of the 24 trials.
So the halo has no, or at most very small, net motion.

To proceed further, we need to look into potential variations with distance and
to whether we can find any more clues regarding the  presence of outliers.
The $v_r(GSR)$ we have determined for \nvr candidate RR Lyr with distances beyond 50~kpc in
the  MW halo are shown as a function of $r$ in Fig.~\ref{figure_vr}.  In this figure,
the stars are divided into three distance regimes, with the intermediate one being 70 to 85~kpc,
and the most distant group, which contains 26 stars, ranges outward from 
85~kpc to 109~kpc, with four at distances exceeding 100~kpc.  
The first point to note is that the mean $v_r(GSR)$
for each of the three groups (shown as large stars in the figure) is close to 0 km/s; the mean values
and other statistics are given in Table~\ref{table_vr_sigma}.  This  is yet another
indication that our values of $v_r(GSR)$ inferred from our $v_r$, as corrected for phase within
the RR Lyr period, are in general  valid. 

Before computing the velocity dispersion, we need to decide whether there are genuine
outliers and how to handle them.  These are important as they
may be a manifestation of previously unknown large scale substructures in the halo.
Fig.~\ref{figure_vr} shows several outliers, 
and we have chosen ${|}{v_r}{|} < 200$ km s$^{-1}$
as the cutoff for outliers for the sample within 85~kpc,
dropping to 170~km s$^{-1}$ outside that distance.  The number of outliers in each
distance range is given in Table~\ref{table_vr_sigma}.  With this definition,
there are a total of 9 outliers from
the \nvr RRab in our Keck $v_r$ sample.  They are circled in Fig.~\ref{figure_vr}.

\begin{figure}
\epsscale{0.6}
\plotone{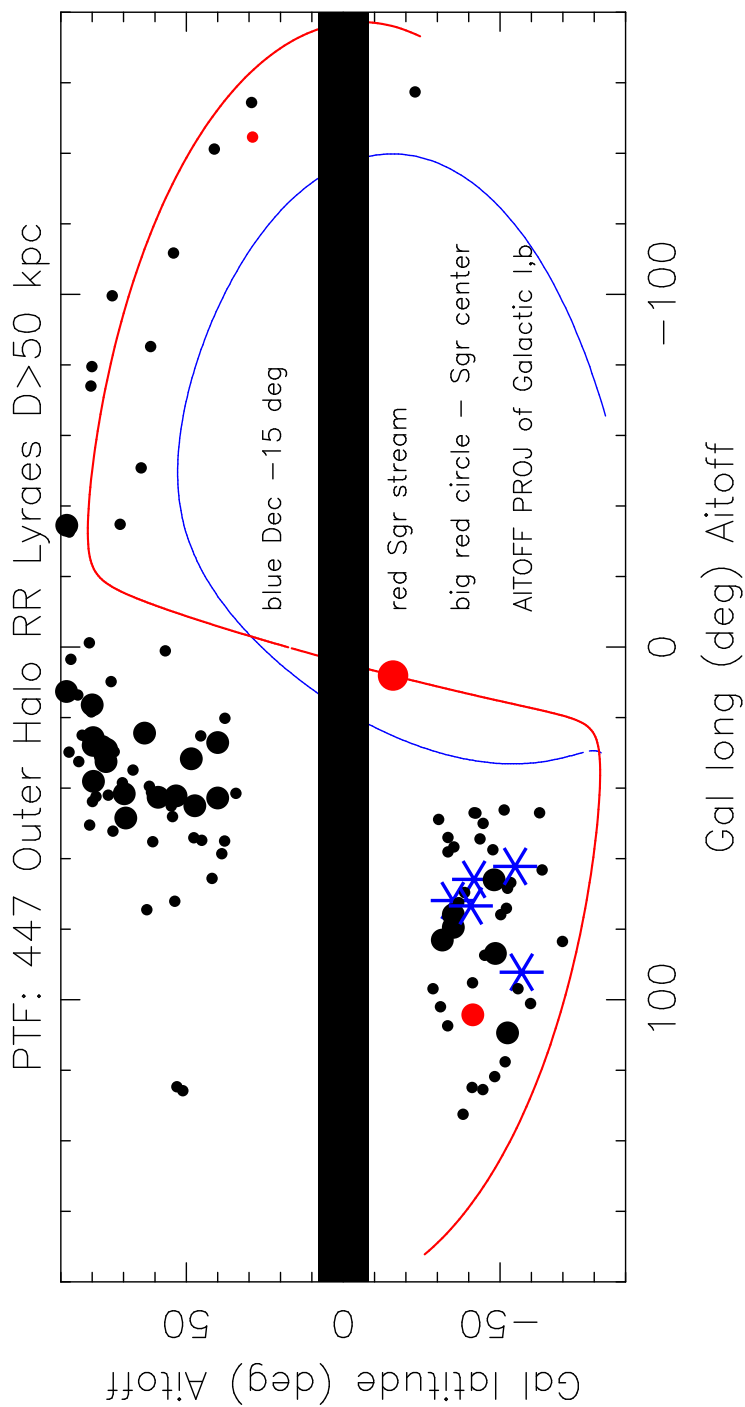}
\caption[]{The location on the sky in galactic coordinates of the PTF  
outer halo RR Lyr stars with  $v_r$ from Keck/Deimos
is shown in an Aitoff projection of Galactic coordinates. 
The larger dots denote RR Lyr with distances beyond 85~kpc. 
 The five large blue stars indicate the only stars of
the sample of 116 which have $v_r(GSR) < -200$~km/sec, while the red points have
$v_r(GSR) > 200$~km/sec.  The area around the Galactic plane that was excluded is indicated
by the solid bar.  The locus of the Sgr stream is denoted by the red curve, with
the nucleus of the Sgr dwarf galaxy indicated by the large red circle.
Dec $-15$~deg is indicated by the blue curve.
\label{figure_radec_vr} }
\end{figure}

\begin{figure}
\epsscale{1.0}
\plotone{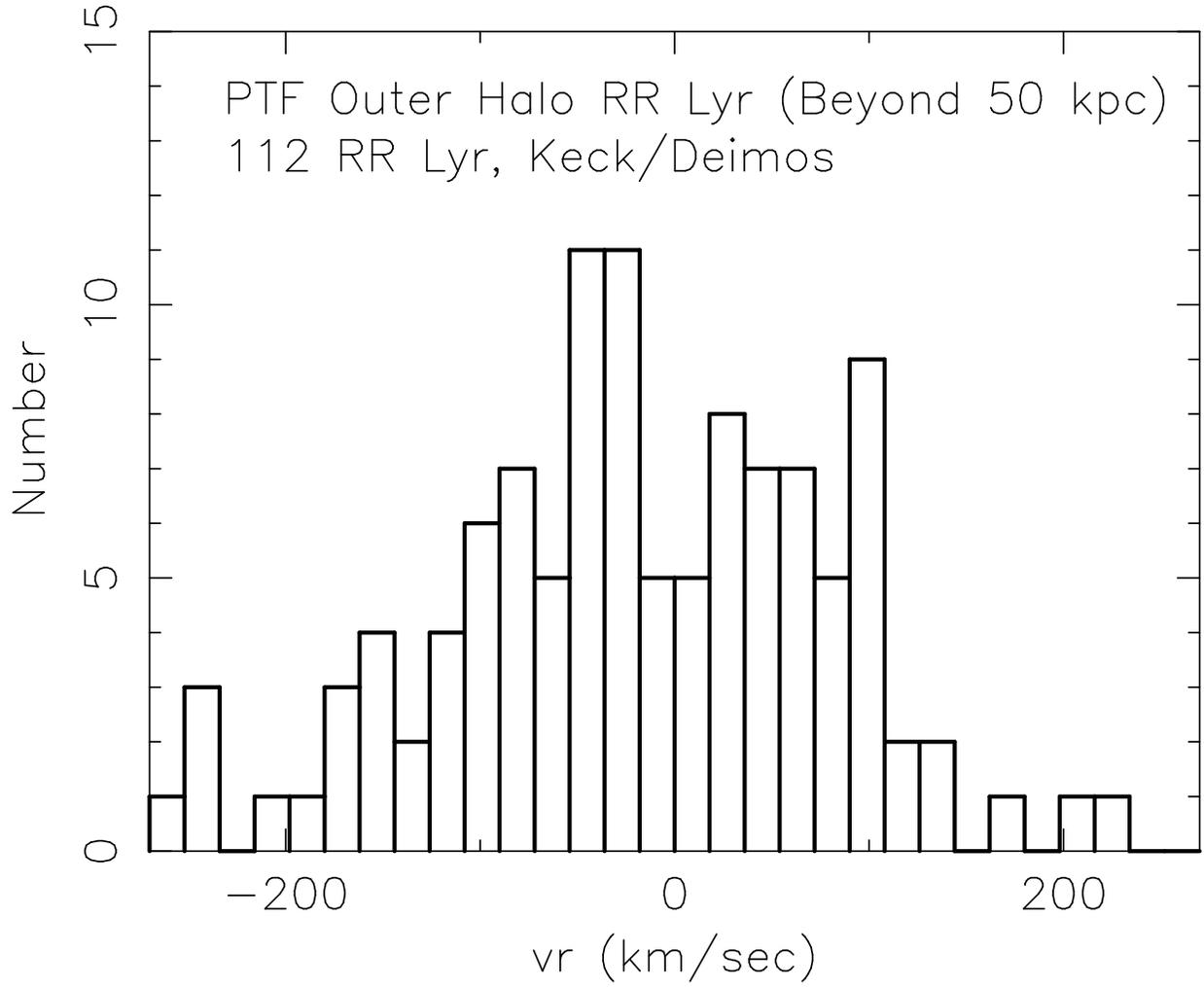}
\caption{A histogram of the Keck/Deimos $v_r$ for the sample of 112 RR Lyr
in the outer halo of the Milky Way beyond 50 kpc.
\label{figure_vr_hist} }
\end{figure}

\begin{figure}
\epsscale{1.0}
\plotone{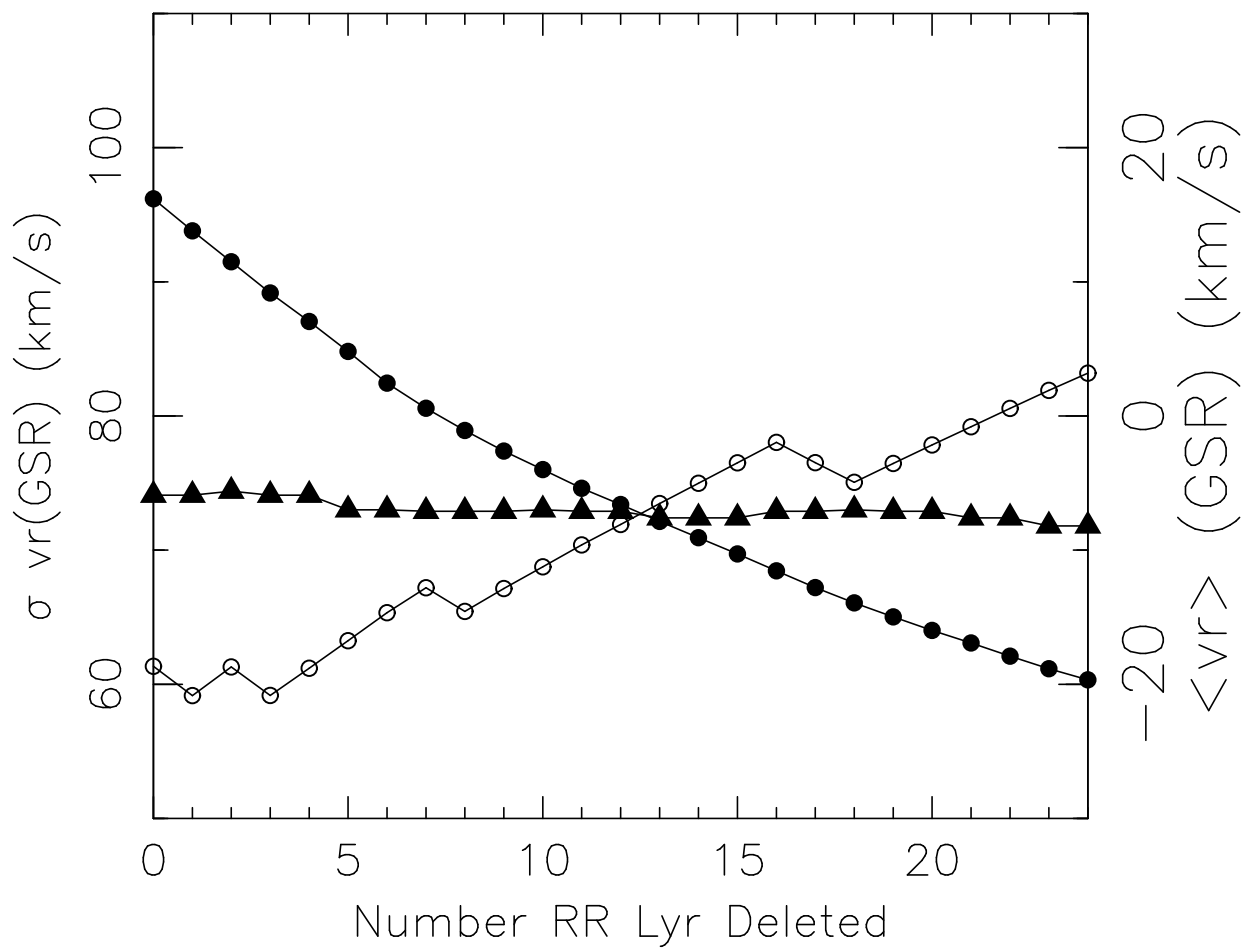}
\caption[]{Tests sequentially deleting the largest outlier in $| (v_r - <v_r>) |$ where
the mean $v_r$ is that of the previous iteration are shown 
for $\sigma(v_r)$ as filled circles, as filled triangles for the median distance,
and as open circles for the median distance of our RRab sample in the outer halo.
The left axis gives the vertical scale for $\sigma(v_r)$, while the right axis
gives the vertical scale for the other two curves.
\label{figure_1x1} }
\end{figure}

\begin{figure}
\epsscale{1.0}
\plotone{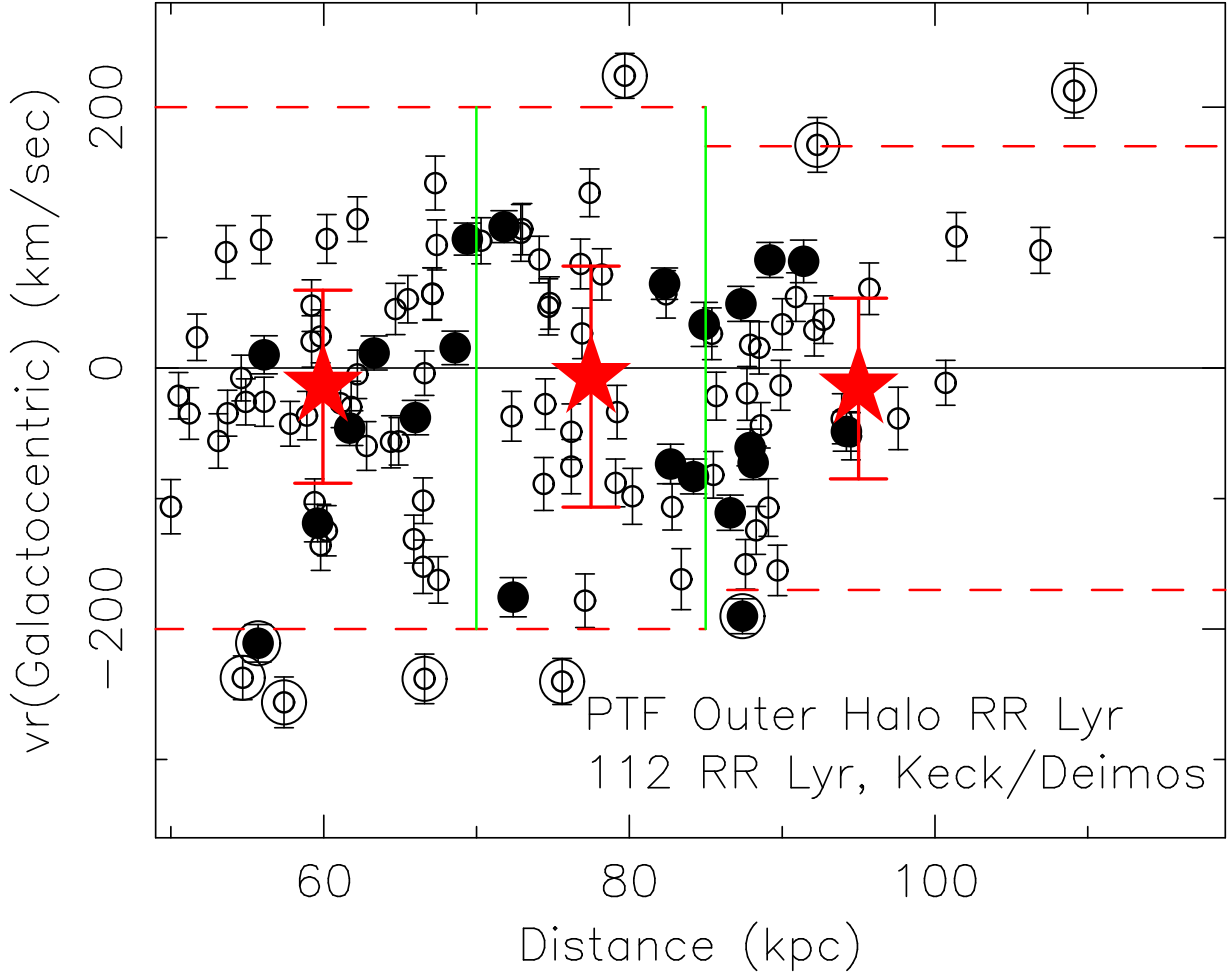}
\caption[]{Radial velocities in the galactic standard of rest are shown
as a function of  distances 
for our sample of 112  RR Lyr selected from the PTF with Keck/Deimos moderate
resolution spectra.
Filled circles denote stars with two Deimos spectra, open circles have 
one spectrum.  $1\sigma$ error bars are shown for each RR Lyr.
The regions (both for high and for low $v_r$) considered outliers in $v_r$
are indicated by the dashed horizontal lines.  The two vertical lines
denote the boundaries between the close, middle, and far samples.
The large stars are located at the median distance for each of these
three samples in X, at the mean $v_r(GSR)$ in Y, and their error bars
indicate the velocity dispersion for each of the three distance groups
ignoring  the outliers.  The 9 outliers are circled.
See Table~\ref{table_vr_sigma} for detailed statistics of the $v_r(GSR)$
distribution.
\label{figure_vr}}
\end{figure}

\begin{figure}
\epsscale{1.0}
\plotone{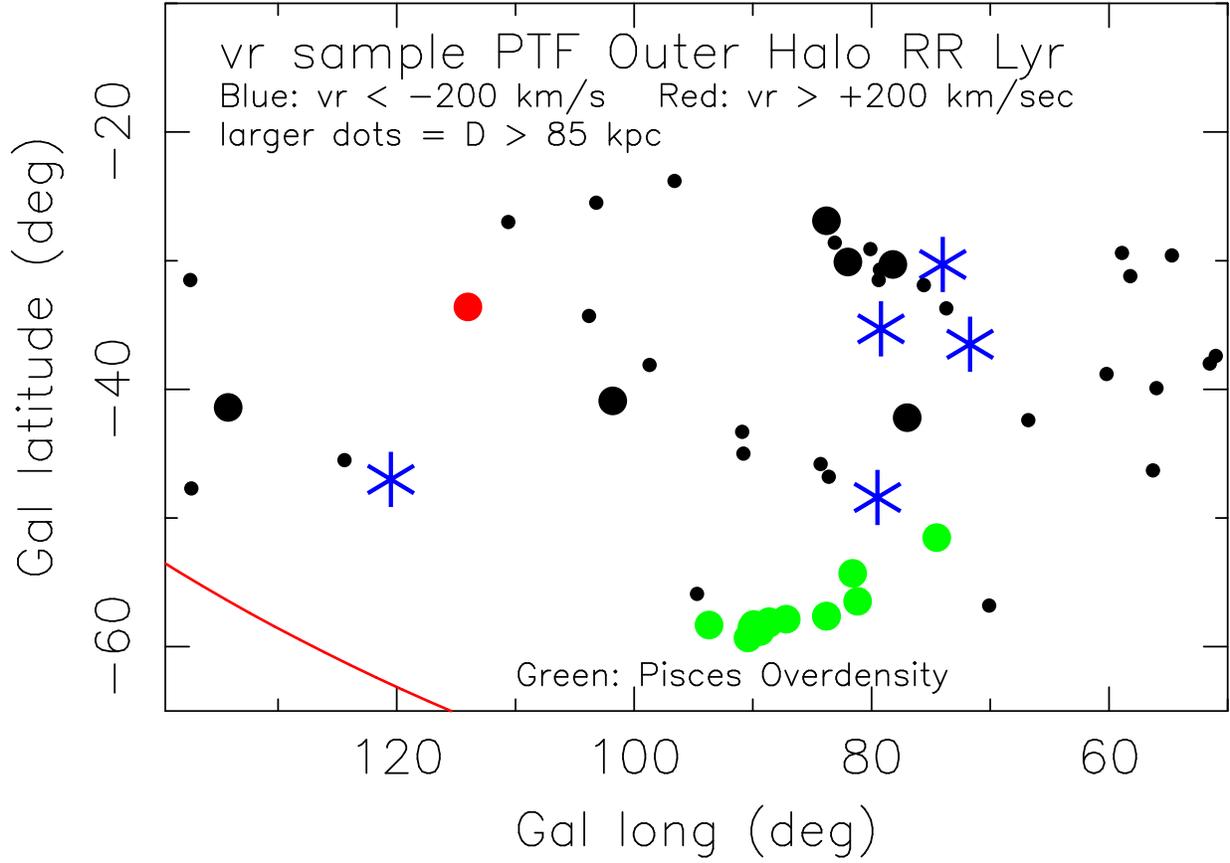}
\caption{The $v_r$ sample is shown in a plot of Galactic coordinates
for a small subset of the total area on the sky covered.  All the candidates
with $v_r < -200$~km/sec, indicated as large blue stars, lie within this small area on the sky.  
The location of the RR Lyr in the
Pisces overdensity found by \cite{sesar07_pisces} is shown in green.  The red curve denotes 
the Sgr stream.
\label{figure_radec_subset} }
\end{figure}

We first look at the low outliers.  There are only five major low outliers.  These five RR Lyr 
have $v_r < -200$~km s$^{-1}$, and distances between 51 and
73~kpc.  All of them have excellent quality 1 light curves.
Fig.~\ref{figure_radec_vr} shows the position on the sky
of the \nvr RR Lyr candidates in our Keck/Deimos $v_r$  sample.  These 5 low
outliers (indicated as
blue stars in the figure) are confined to a small
region on the sky with RA between 329 and 10$^{\circ}$ and Dec between 4 and 15$^{\circ}$.
A blowup of this region on the sky is 
shown in Fig.~\ref{figure_radec_subset}.  The
green points indicate RR Lyr in the Pisces overdensity, discovered
by \cite{sesar07_pisces} as a linear stream
at a distance of $\sim$80~kpc within the SDSS stripe 82; however recall that  Stripe 82
is a narrow equatorial stripe only 1$^{\circ}$.27  wide
extending from R.A.~20$^h$ to R.A.~4$^h$.  A more recent discussion of
this structure is  given by \cite{nie_pisces}, but this structure is more distant
than the set of 5 low outliers in our sample.
An examination of Figs.~\ref{figure_radec_vr} and \ref{figure_radec_subset}  combined with
Fig.~\ref{figure_vr} strongly suggests that these five stars belong to some previously unknown
diffuse outer halo structure
which extends over about 40$^{\circ}$, perhaps from a disrupted satellite.
 These five low outliers are sufficiently far from the plane of the
Sgr streams that it is unlikely that they are part of it.
We therefore consider them as potentially not virialized. 
The choice of the $v_r$ cutoff adopted for RR Lyr with $r > 85$~kpc 
of ~ ${|}v_r{|} < 170$~km s$^{-1}$ is an estimate
based on Figure~\ref{figure_vr}.  

Table~\ref{table_vr_sigma} gives the statistics of the sample of \nvr RR Lyr
when divided into three  distance ranges, and when only two groups are used, with a boundary
at 85~kpc.  Values for $\sigma[v_r(GSR)]$ are given as calculated from the measured
$v_r(GSR)$ of each RR Lyr, and also with a 20~km s$^{-1}$ measurement error removed.
They are calculated for the full sample, and also for the case where the major
outliers have been excluded.  Since the removal of only a few outliers
considerably reduces the velocity dispersion within each distance range, we consider our choice
of cutoffs for outliers as reasonable.  For example, for the outermost group
of  RR Lyr
with $r > 85$~kpc, $\sigma[v_r(GSR)]$
is reduced from 90~km s$^{-1}$ to 65~km s$^{-1}$ by removing only 3 outliers from our sample of
26 RR Lyr in this distance range.

Note that with the outliers eliminated, the velocity disperion is quite low, not exceeding
87~km s$^{-1}$ beyond 50~kpc, and for the outermost stars with $r > 85$~kpc, 
$\sigma[v_r(GSR)] \sim 65$~km s$^{-1}$. Clearly a larger sample 
of tracers with accurate distances and $v_r$ beyond 50 kpc is desirable.  We
are working on it, but it will take several years to enlarge our
sample of \nvr RRab beyond 50~kpc with measured $v_r$ by a substantial factor.

\section{The Radial Distribution of the PTF RR Lyr Sample \label{section_nradial} }

With considerable caveats, we present the radial distribution in the outer
halo of the MW for our sample of \ncand RR Lyr candidates.  We assume an
isotropic spherical halo.  The major concern is the serious incompleteness in our sample
of outer halo RR Lyr stars at the largest distances probed, i.e. beyond 90~kpc.

One might also worry about an increasing number of interlopers (i.e. not genuine
RRab) in the sample as the distance increases and the light curve quality
decreases due to increasing observational uncertainties in each individual observation
and to a lower fraction of detections coupled with an increasing fraction of upper limits
for a given number of epochs of observation.  However, the small spread of the
period distribution and the strong period-amplitude correlation shown in
Figs.~\ref{figure_period_hist} and \ref{figure_period_amp}
as well as the behavior of the quartiles of period and amplitude as a function
of distance (see Fig.~\ref{figure_quartiles_per_amp}        suggest that our
sample is not contaminated by interlopers even at the largest distances we probe.
Note that a careful examination of Fig.~\ref{figure_quartiles_per_amp} does
support the suggestion that there is a strong increase in incompleteness of our sample
of RRLyr at the
largest distances included our sample. 

We can assess the importance of incompleteness by considering the fraction of upper limits
instead of detections among RRab
which have many epochs of observation
and which span the full range in distance probed here.
Fig.~\ref{figure_ul} shows some relevant data, specifically the fraction of upper limits
among the available images in the PFS database
for a sample 80 RR Lyr stars at the close end of our sample ($\sim$50~kpc) vs 50
of the most distant ones (i.e. beyond 95~kpc).  
As we had no control over
the cadence nor of the selection of fields to be observed on a given night at that time,
the number of observations of a given field 
(at least at the time that the sample was constructed) depended on
how many times a field was observed by other PTF projects.

For the nearer RR Lyr in our sample, the fraction of upper limits
is low, usually less than 10\%, while for the most distant ones, the fraction
of upper limits is typically $\sim$60\%.  Our   PTF RR Lyr sample was selected in 
2014 and hence given the much smaller number of images of each field in the database
at that time compared to the present values given in Table~\ref{table_vr},
a larger fraction of the most distant RR Lyr will not be picked up as candidate RR Lyr
as their light curves would not have contained more than 30 detections at that time. 
Although as indicated earlier in \S\ref{section_light_curve_quality}, 
the ephemerides have been checked recently and updated 
as necessary, the list
of candidates has not.  Redoing the selection from the PTF seems unjustified given that 
a high quality PS1 RR Lyr
catalog with careful determination of its completeness and purity will be released shortly
\citep{nina_ps1,sesar_ps1}.
We can safely assume that there is an incompleteness of at least
a factor of two for the most distant part of our sample; the actual completeness correction
at the faint end of our PTF sample could be even larger.

With regard to the issue of   contamination of the Keck $v_r$ sample by interlopers, we compare the
number of candidate RR Lyr with $Pr > 0.8$ as a function of distance 
with the distribution in distance of our sample of \nvr RR Lyr with Keck/Deimos $v_r$.
Note that 326 of the \ncand candidates meet this probability
restriction, while essentially all of candidates selected for Keck spectroscopy 
have $Pr > 0.8$.
This ratio is given as a function of distance
in Table~\ref{table_frac_vr}.   There are many candidates at the near end of the sample,
but our goal was to get spectra of as many distant stars as possible, so
 candidates at the closer end
of our sample were not observed unless no suitable distant candidate had a phase
within the allowed range during that part of the night.  
As shown in Table~\ref{table_frac_vr}, only 21\% of the candidates with $r < 64$~kpc
have Keck $v_r$, while this fraction is $\sim$50\% from 64 to 99~kpc,
beyond which it drops to $\sim$38\%.
If the sample of candidates was seriously contaminated with
interlopers as the distance increased, this success fraction should have fallen significantly.
We can therefore assume that the fraction of interlopers is not rising significantly
towards the faint end of the sample, until a distance of at least 90~kpc, beyond which
the sample is  small.

Another way to approach the same issue is to examine the current PTF light curves for
those RR Lyr candidates which are not in the Keck $v_r$ sample.  If there is serious contamination
which is dependent on distance
by stars which are not RRab 
(presumably more contamination at larger distances), the fraction of these RR Lyr candidates
that have quality 3 light curves (light curves which do not suggest that the star is a RR Lyr)
 will rise substantially with distance.  We have carried out this check
for candidates over a wide range in distance.  For candidates at distances closer than 60~kpc,
94\% of them show high quality light curves.  This fraction falls to 85\% for
those between 65 and 70~kpc.  It remains above 80\% out to 95~kpc, beyond which
it drops to $\sim$70\%.  This change with distance of the potential fraction of contaminants
is small enough that the effect on the power law fit should not be large, at least
within 95~kpc. 

Fig.~\ref{figure_distance_hist} presents the number of candidates in bins in distance,
with both axes of the plot using a logarithmic scale.  Also shown are a number of
power law fits.  The upward arrow indicates a  correction for an incompleteness
of 50\% in the
RR Lyr sample at the
largest distances probed arising from the large fraction of upper limits in their PTF light curves.
It is clear that a volume density law of $\rho(r) \propto r^{-4}$ is a reasonable fit 
from 50 to 85~kpc, and, with an incompleteness correction of a factor of two,
 would be a good fit  out to 100~kpc.

\section{Comparison With Previous Results \label{section_compare} }  

\subsection{Comparison With Other Samples of Milky Way Outer Halo Stars
\label{section_other_samples} }

As a result of many recent large stellar surveys,
our knowledge  of the outer halo of the MW is improving very rapidly.
Outer halo stellar samples are increasing in size and distance
range probed.
For comparison, the early discussion of the kinematics of the halo 
by \cite{kinman96} used only a total of
67 RR Lyr and BHB stars in the inner halo out to $r < 15$~kpc;
they  found a velocity dispersion of $\sim$110~\kms.  However, samples
of outer halo stars with spectroscopic $v_r$, particularly in
the crucial region beyond $r = 80$~kpc, such as ours are growing very slowly.

The Catalina Real-time Transient Survey (CRTS) 
has been in operation since 2006 mining
the data stream
from three telescopes (0.7 m, 1.0~m, and 1.5 m diameters) in the mountains
north of Tucson  Arizona
which are operated by the Lunar and Planetary Laboratory at the University
of Arizona and
whose primary mission is the detection of near earth asteroids.  
The photometric
calibration and cadence of the CRTS are not as well controlled as those of the PTF,
but the time span of the imaging and hence of the light curves is more than a decade.
\cite{drake_crts} used this database to produce the Catalina
Surveys Periodic Variable Star Catalog, which has roughly 16,800 RRab 
variables.   The maximum distance
of the RRab in their survey is $\sim$60~kpc, however the bulk
of their sample is closer than 40~kpc.  Our sample begins at 50~kpc
and  we have eliminated
the Sgr stream, while the CRTS catalog has not. 
We find that the overlap between our sample of \ncand\ RR Lyr
and their sample is only 32 stars. The agreement of the derived period for the variables
in common between the two surveys is less than 0.0010 days for 24 of the 
32 stars in common, while the largest difference is 0.0016 days.
The more recent southern extension of the Catalina Sky Surveys RR Lyr
catalog, \cite{torrealba_crts}, has no overlap in sky
coverage with our Palomar based survey. 

Very recently \cite{gaia_rrlyr}, in a paper not yet accepted,
have produced a catalog of RR Lyr by
combining the first GAIA data release \citep{gaia_dr1} with 2MASS \citep{2mass}.
Their sample has $\sim$21,600 RR Lyr and is confined to the inner halo.
Thus there is a very large overlap with the CRTS
sample of \cite{drake_crts}, but \cite{gaia_rrlyr}
only reach out to a heliocentric distance of 20~kpc, and thus
there is no overlap with our sample, whose minimum distance
is 50~kpc.

The huge database of SDSS, coupling uniformly measured multi-color photometry from
its deep imaging over a large fraction of the northern sky and uniformly reduced spectra, 
 was a breakthrough.  It was used by \cite{xue08} to isolate
a sample of $\sim$2400 BHB stars with $v_r$ that reaches out to $r \sim 40$~kpc, with
very limited coverage out to 50~kpc; \cite{xue11} gives a slightly improved selection
of BHB stars from the same material.
More recently, \cite{xue14} used the database of the
Sloan Extension for Galactic Understanding and Exploration (SEGUE) \citep{segue}
to select a sample of 6036 distant K giants.
They developed
probabilistic procedures to obtain their luminosities, claiming to thus have achieved a median
accuracy of 16\%  in their distances. 
Their sample extends out to $\sim$80~kpc, although almost all of the stars beyond 60~kpc are
in the Sgr Stream or other known halo substructures.

The K-giant sample of \cite{xue14}, when cleaned of known substructures, primarily the 
Sgr Stream, has  1757 stars with distances beyond 10~kpc, but in the outer halo it is 
significantly smaller than our sample, which begins at 50 kpc.  The purged sample
of \cite{xue14} has only
two K giants beyond 65~kpc, with the most distant at about 75~kpc.
Our sample 
reaches significantly further out in the MW halo
with a median distance of 73~kpc.  Furthermore, the distances to our RR Lyr sample are much
more accurate than those of K giants.

The hypervelocity star survey \citep{hvs_2014} has carried out extensive spectroscopy of very blue
stars in the outer halo selected from SDSS photometry.
While SDSS colors are used, this is one of the few surveys besides ours that
obtains their own spectra.  The kinematics of the majority of their sample 
of the late B-type outer halo stars
found the course of this work are discussed in \cite{brown_hvs}.  Their dataset 
contains 910
late B and early A such stars, almost all of which are BHB stars
with a small contamination of less luminous blue stragglers.   The
bulk of their sample is closer than 50~kpc.

\cite{bochanski14} selected a sample of 404 candidate very distant M giants based on their 
NIR colors from UKIDSS combined with optical colors from SDSS and undetectable proper motions
(to rule out nearby M dwarfs).
Two of these were spectroscopically confirmed and appear to be extremely distant,
with their estimated minimum distances being 130 kpc.  Their sample of very distant M giants
selected via photometry has roughly 80\% contamination which can only be resolved by spectroscopy;
photometry alone is insufficient.  Furthermore, M giants in the outer halo are a very biased
indicator as they can
only arise from a metal-rich population, and  presumably are located in 
potentially non-virialized initially compact
infalling structures, if in fact their distances and classifications are correct.
As noted by \cite{bochanski14}, these stars lie close to the Sgr plane.
The recent model of the Sgr Stream by \cite{extended_sgr} suggests these M giants
are located within the Sgr Stream; it successfully reproduces their
distance and low $v(GSR)$. \cite{sesar_sgr_spur} have recently identified some of these
spurs in the Sgr stream at distances exceeding 100~kpc using the PS1 RR Lyr sample.
However, given the high contamination fraction of their M giant sample,
the amount of observing time which would be required to
generate a clean large sample of such distant M giants is prohibitive and
furthermore a sample of M giants would not probe the bulk of the outer halo of the MW.

\cite{slater17} isolate a sample of  $\sim$4000 distant giants in the halo 
with wide field imaging using a narrow bandwidth
filter covering the region of the Mg triplet at 5170~\AA, which
is well known to be a good giant/dwarf discriminator. 
This is combined with broad-band SDSS
imaging.     Extensive statistical treatment
using population synthesis  modeling  is required to clean the sample
of numerous dwarf interlopers, and the distances of individual objects are quite uncertain.
The sample of $\sim$4000 giants reaches out to 80~kpc.

\cite{deason_veil} attempted to build up a sample of more distant ($D > 80$~kpc) BHB stars by stacking
multi-epoch photometry from Stripe 82 of the SDSS (and other regions with multiple
images) to isolate candidate BHB stars, 
but these  are too faint to have SDSS spectroscopy.  They  obtained low resolution 
($\lambda / \Delta\lambda \sim 800$) spectra using FORS2
on the VLT to try to separate BHB stars from contamination by brighter blue stragglers,
which outnumber by a factor of more than four the desired BHB stars.  The final sample has only 7
faint BHB stars.  \cite{deason_veil} then add a small number
of other potential outer halo stars with highly uncertain distances as well as the dwarf satellites
of the MW.

The only sample that reaches out to the distances probed by our RR Lyr sample 
with a substantial number of stars beyond 50~kpc is that of
\cite{deason14} which uses BHB and blue straggler stars from the SDSS DR9.  There are several issues
that afflict this sample, particularly
contamination with blue stragglers, and, more seriously, with QSOs.  Extensive color
modeling, taking into account photometric scattering, was used to try to remove contaminants, which
outnumber the desired BHB stars by a large factor.

It is clear that the sample of distant RR Lyr from the PTF with Keck radial velocities presented here 
has unique characteristics. It is a clean sample with few interlopers, and
each star has has a highly accurate distance. At the present time, and even after the
release in Nov. 2017 of the PS1 sample by \cite{sesar_ps1},
ours is the only  reliable sample with at least a modest number of of tracer stars beyond 80~kpc
in the outer halo with measured $v_r$.

\subsection{The Density Profile in the Outer Halo \label{section_density} }

The determination of the density profile in the outer halo of a large
set of massless tracer stars is clearly a crucial input to determining
the mass of the MW. Given the limited data, the solution is
usually expressed as a power-law fit to the density vs distance.
Our preliminary result based on a large sample of RR Lyr variables 
is given in \S\ref{section_nradial} and is shown in Fig.~\ref{figure_distance_hist}.
We find 
that a power law in $r$ with a slope of $\sim -4$ is consistent with
the stellar density $\rho(r)$ derived from 
the distances of our RR Lyr sample.  This assumes an isotropic spherical halo.
With  larger samples one can also solve for the flattening profile of the halo,
but we could not attempt this.  \cite{bovy17}, based on an analysis
of the Pal 5 and GD-1 stellar streams, suggest that the axis ratio of the dark matter's
halo density distribution is 1.05 within the inner 20~kpc, providing some
support to our assumptions, although \cite{gaia_rrlyr} suggest that the inner
halo has a substantial oblateness which decreases at larger Galactic radii. 
There seems to be a general consensus that
the outer halo is less oblate than the inner halo.

Our result contradicts that of \cite{deason14}, who claim that beyond 50~kpc
there is a striking drop in the stellar halo density.
Although in their earlier paper \citep{deason11} they found a power law fit
of $\alpha = -4.6$ for the region $27 < r < 40$~kpc (the maximum $r$ reached),
\cite{deason14} find a power law slope of
$-6$ beyond 50~kpc, with even steeper slopes
(power law index $-6$ to $-10$) favored at larger radii.  
On the other hand, \cite{depropris10}, who
used a sample of 666 BHB stars from the 2dF quasar redshift survey, found a
very shallow slope for the density in the outer halo 
of $-2.5{\pm}0.2$ and a velocity dispersion which increases with $r$,
reaching a huge $\sigma(v_r)$ exceeding 200 km s$^{-1}$ at $r \sim 80$~kpc
over the two lines of sight probed.  Our data do not
support the results of either of these two studies. 
Our sample is much cleaner with much better distances than the samples of either
of these two analyses.

A number of other analyses have been published recently which agree with
our halo density distribution to within the uncertainties.
Among the many samples of outer halo stars
discussed above in \S\ref{section_other_samples}, the large SDSS/SEGUE
samples of K giants stand out for their size and spatial coverage.  The latest analysis of such
is that of \cite{xue16}.  As the luminosity of K giants depends strongly on the
metallicity, they had to use forward modeling techniques to fit the spatial distribution
and abundance distribution simultaneously.
They found that a power-law slope with index $-3.8\pm 0.1$ is a good fit 
to the number density profile of the halo beyond $r \sim 20$~kpc.
\cite{das16} reanalyze this sample 
using an extended distribution function  to find the density distribution
power law index is $-4$ at large radii out to 80~kpc.
\cite{kafle14}  combine the BHB and K giant samples from the SDSS/SEGUE to find
a slope of $-4.5$ in the halo beyond $\sim$20~kpc.
The recent work by \cite{slater17} using SDSS photometry coupled
with imaging in a narrow band filter centered at the Mg triplet to
eliminate dwarfs also targets K giants. They use sophisticated CMD modeling
and population synthesis to derive
a halo density profile $\rho \propto r^{-3.5}$ from 30 to 80~kpc.

Thus, as discussed above, there seems to be a growing consensus that in the 
outer halo of the MW at least out
to 85~kpc, the stellar density can be represented as a power law with a slope
of $-3.5$ to $-4$.  This is quite close to the slope found in the
inner halo, at least from 20~kpc outward, by several groups, see e.g. \cite{xue16}.

\subsection{The Velocity Dispersion In the Outer Halo}

\begin{figure}
\epsscale{1.0}
\plotone{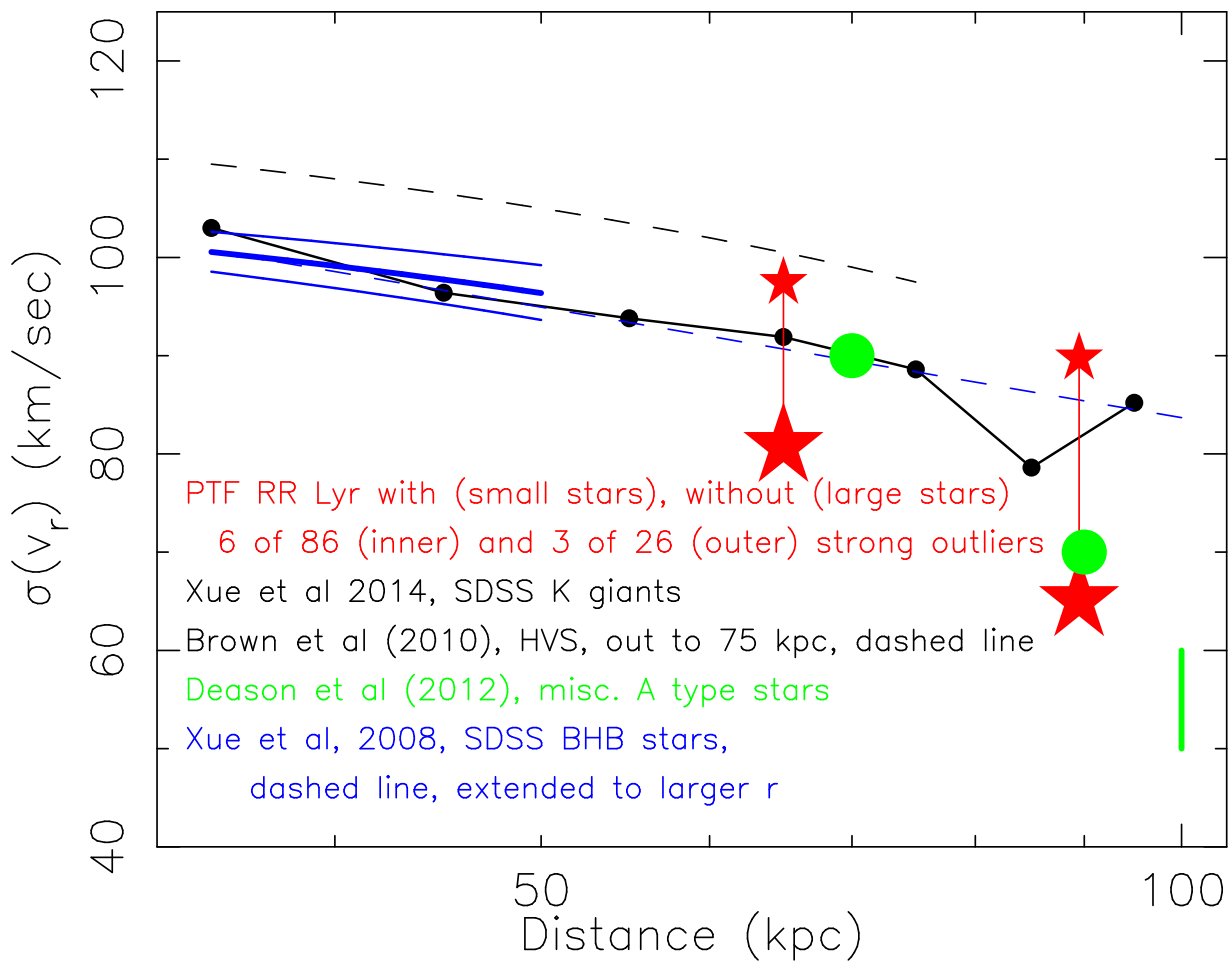}
\caption[]{$\sigma(v_r)$ in the Galactocentric rest frame are shown  for
our inner and outer sample of RRab from the PTF (split at 85~kpc), with (large red stars) and 
without (smaller red stars)
eliminating the strong outliers (3 in the outer sample, and 8 in the inner sample).
Values that have been derived in several recent
studies by \cite{xue08}, whose extrapolation to larger $r$ is indicated
by a dashed line, \cite{xue14}, \cite{hvs_2014}, and \cite{deason_veil}
are also indicated.  Note that the X-axis has a logarithmic scale.
\label{figure_allsigma} }
\end{figure}

\begin{figure}
\epsscale{1.0}
\plotone{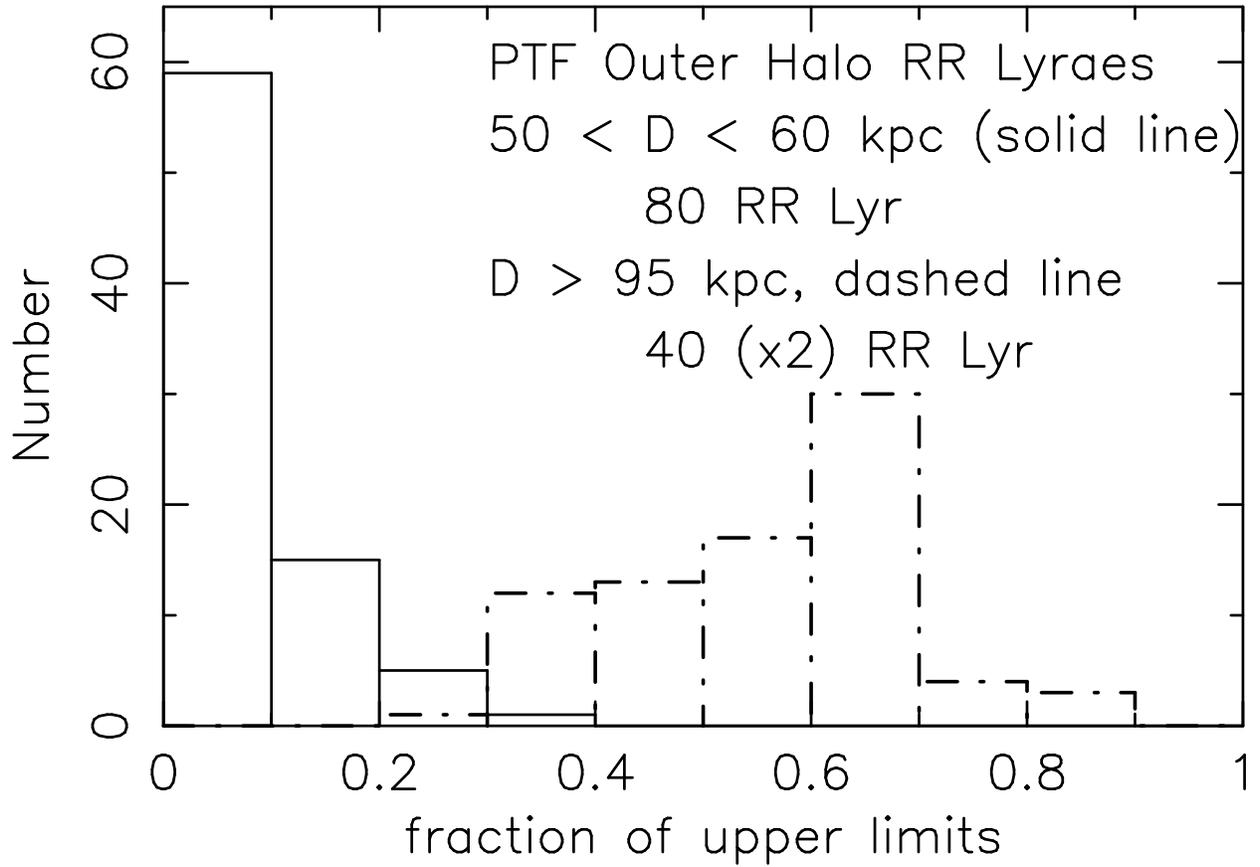}
\caption{Fraction of upper limits among the total observed epochs
of the PTF are shown for samples of candidate RR Lyr with distances $\sim$55 kpc vs those
with distances $>$~95 kpc. For the most distant RRab, a much larger fraction of the
PTF images do not result in a detection of the candidate RR Lyr.
\label{figure_ul} }
\end{figure}

\begin{figure}
\epsscale{1.0}
\plotone{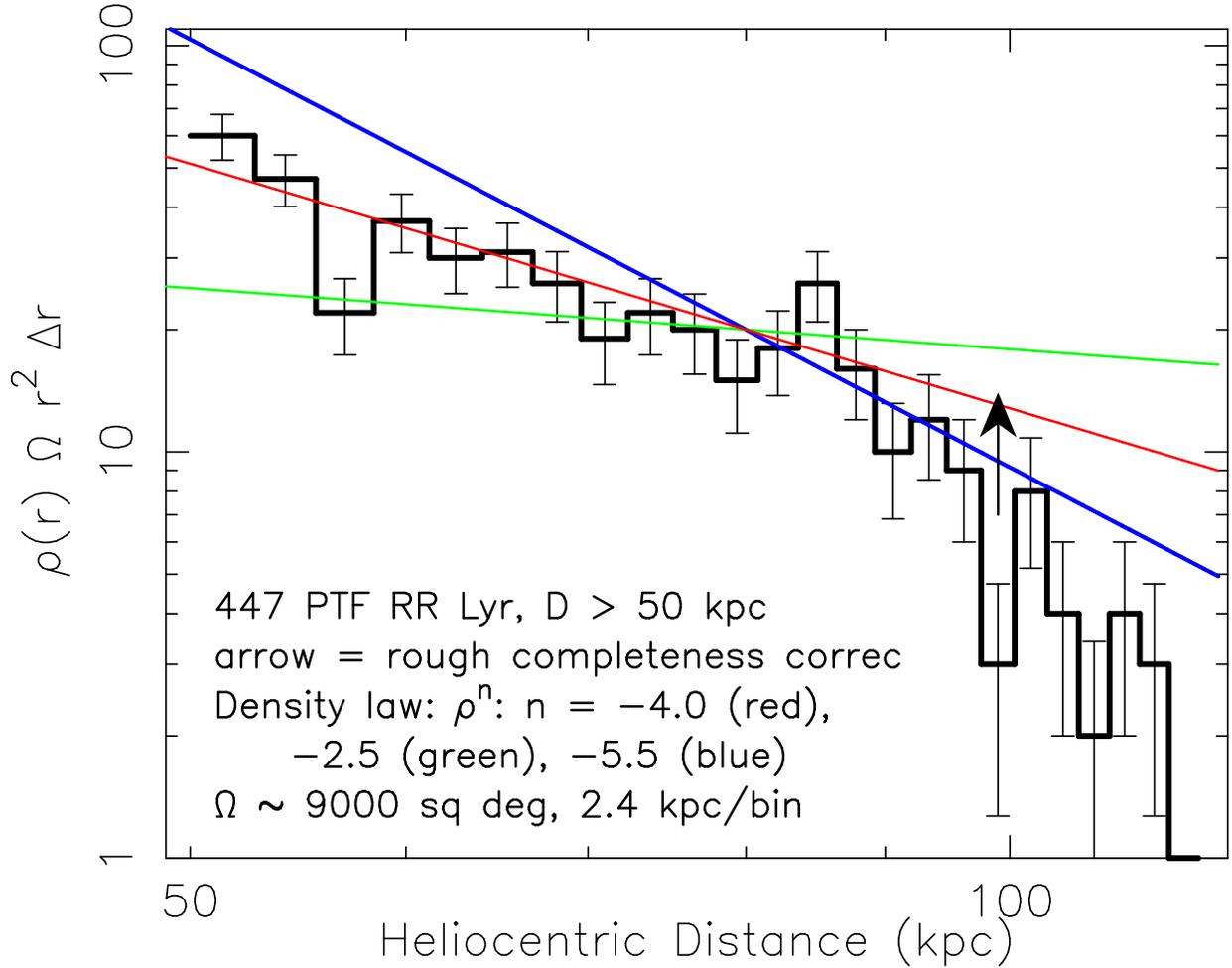}
\caption{Histogram with distance of the 447 candidate RR Lyr stars
from the PTF sample beyond 50~kpc.  Power laws for $n ~ =  ~ -2.5$~,~ $-4.0$, and    
$-$5.5 are shown.  The data are reasonably well fit for $n \sim -4$ out to ${\sim}$90~kpc,
after which a steeper slope is seen.  However at that distance, the incompleteness
effects are severe; a guess at the minimum completeness correction at such
distances is shown by the upward pointing arrow.
\label{figure_distance_hist} }
\end{figure}

The behavior
of the $v_r$ of a sample of massless tracers  as a function of distance 
provides important clues as to the potential and
total mass of the MW.  Towards this goal, several of the studies referenced in 
\S\ref{section_other_samples} have measured
$v_r$ for a large fraction of the members of their sample.  In particular
those based on SDSS and its successors (i.e SEGUE) fall into this class.
In this section we compare our derived $\sigma(v_r)$ as a function
of distance for our RRab sample (shown in Table~\ref{table_vr_sigma}
and in Fig.~\ref{figure_allsigma}) with those of other groups.

The two large samples of outer halo stars based on the SDSS and SEGUE, i.e. 
the BHB sample of \cite{xue08} \citep[see also][]{xue11} and the K giant sample of \cite{xue16},
both of which reach out to $r \sim 50$~kpc,
have been analyzed by many different groups using various 
sophisticated modeling techniques to derive properties
of the outer halo.
The latest result from these samples is \cite{xue16}, where references to earlier
work can be found.

\cite{xue08} and \cite{xue11} derived the radial trend of $\sigma(v_r)$ 
out to 50~kpc, where they found $\sigma(v_r) \sim$95~km s$^{-1}$.  The spatial range of this relation
was extended by \cite{deason_veil}, who added a small number of more distant objects.
\cite{kafle14}, who derived their own sample of K giants from the SEGUE data, also
found a similar value of $\sigma(v_r)$ of $\sim$100 km s$^{-1}$
for $r \sim 50$~kpc; see their Fig.~1.

Fig.~\ref{figure_allsigma} illustrates some of these results from the literature
compared to our relationship for $\sigma(v_r)$ as a function of $r$ between
50 and 100~kpc.  With the exception of  \cite{depropris10},
all of these investigations, including ours presented here,
are in reasonable agreement regarding the velocity dispersion 
of the outer halo stars as a function
of distance from 50 to 100~kpc within the regime probed by each group, 50 to 100~kpc
in our case.  
All recent studies find $\sigma(v_r)$
$\sim$90~km s$^{-1}$ at 50~kpc, dropping lower as $r$ increases.
The hypervelocity star survey \citep{hvs_2014}
derives the same general decline of $\sigma(v_r)$ with $r$ but has $\sigma(v_r)$ roughly 20~km/sec higher
at all $r$ probed than our result and that of most recent work.

The agreement on the spatial distribution $n(r)$ among the various studies, among
the most recent of which is \cite{xue16}, 
is also satisfactory out to perhaps 60~kpc;
from $\sim$30 to $\sim$60~kpc all groups agree that the number density of tracers
can be represented by a power law with index of about $-4$.
There are only two surveys beyond that, our work and that of \cite{deason14},
and there is a major disagreement at these larger distances between us, with 
\cite{deason_veil} claiming a very rapid drop in the number density beyond 50~kpc.
They find a power law of $-6$ with distance beyond 50~kpc, dropping to slopes
of $-6$ to $-10$ at larger distances.  Unless we have badly underestimated
our contamination problems, which at least in the sample selected for
Keck spectroscopy is highly unlikely given the quality ratings of the light curves
and the period-phase relation for our sample shown in Fig.~\ref{figure_period_amp},
we advocate that our results are more reliable, given the substantial contamination
of the \cite{deason_veil} sample by QSOs for which the corrections they use may not
be adequate.  

We thus conclude that the outer halo at $r > 70$~kpc is cold,  and it's 
radial velocity dispersion is low.
These factors suggest, in accordance with several recent analyses,
a low total mass for the MW.  For our  RRab
survey based on the PTF database, the key issues are the purity and completeness  of the sample and the
potential impact of substructure which we suggest may produce the
outliers in $v_r$ in these distant outer halo samples of 
``massless and virialized'' tracer stars.
The new PanSTARRS RR Lyr catalog by \cite{nina_ps1} and by \cite{sesar_ps1}
will allow future investigations to avoid most if not all of these concerns.

\section{Summary}

RR Lyr stars of type ab are ideal massless tracers that can be used to study the
outer halo of the MW.  Because they have (to first order) a fixed
luminosity, their periods are about 0.5 days, they are common in old
metal-poor stellar populations, and their amplitude of variation is
substantial, reaching up to $\sim$1 mag, they are easily found in
any multi-color imaging survey with extensive temporal coverage.
Since they are blue, even when they are in the outer halo of the MW, 
they  stand out against the numerous  redder foreground stars, 
and can be distinguished from quasars by the
nature of their variability, quasars being non-periodic variables,
while the other blue halo stars (BHBs and blue stragglers) can be eliminated as being non-variable.
RR Lyr are thus ideal probes of the outer halo which can be found
at great distances in the current generation of large
stellar surveys and whose distances can be measured to high accuracy with
just a light curve.

We present here a sample of \nvr RRab beyond 50 kpc in the outer halo of the
MW for which we have obtained moderate resolution spectra with
Deimos on the Keck 2 Telescope.  Four of these have distances exceeding 100 kpc.
These were selected from a much larger
set of \ncand candidate RR Lyr
which were datamined using machine learning techniques
applied to the light curves of variable stars in the Palomar Transient Facility 
database.  
The observed radial velocities taken at the phase of the variable
corresponding to the time of observation were converted to systemic radial
velocities in the Galactic standard of rest.  This  only works well when
the ephemerides of the variable stars are accurately known.

From our sample of \nvr RR Lyr with Keck $v_r$ we determine the radial velocity
dispersion in the outer halo of the MW to be $\sim$90 km s$^{-1}$ 
at 50~kpc falling to about 65 km s$^{-1}$ near 100~kpc once a small number of major outliers
are removed.  The five very low $v_r(GSR)$ stars, all of which have $v_r(GSR) < -200$ km s$^{-1}$,
are surprisingly close together on the sky at a distance of about 60~kpc, but there
is no known structure at that distance in that part of the sky.

With reasonable estimates of the completeness of our sample of \ncand
candidates and assuming a spherical
halo, we find that
the stellar density in the outer halo declines as $r^{-4}$.  Most, but not all, other recent
works corroborate this functional form.

The problems we have faced have been in the accuracy of the ephemerides for the RR Lyr sample
and in issues of completeness and non-RR Lyr interlopers.  Further exploration 
of the  issue of substructure
in the outer halo requires a larger sample.
The new Pan-STARRS RR Lyr catalog by \cite{nina_ps1} and by \cite{sesar_ps1} provides this,
and
will allow investigations which we expect to carry out
in the near future to avoid most if not all of these concerns.  Ultimately LSST
will allow techniques similar to those we used to identify RR Lyr at even larger
distances of up to several hundred kpc.  Of course spectroscopic follow up of
the very distant RRab we expect to find with  LSST will require the next
generation of extremely large telescopes beyond the current 10~m Kecks.

\acknowledgements

We are grateful to the many people who have worked to make the Keck
Telescope and its instruments a reality and to operate and maintain
the Keck Observatory.  The authors wish to extend special thanks to
those of Hawaiian ancestry on whose sacred mountain we are privileged
to be guests.  Without their generous hospitality, none of the
observations presented herein would have been possible.

We thank the referee for helpful detailed comments that improved this paper.

This work uses data obtained with the 1.2-m Samuel Oschin
Telescope  at  Palomar  Observatory  as  part  of  the  Palomar
Transient Factory project, a scientific collaboration among the
California Institute of Technology, Columbia University, Las
Cumbres Observatory, the Lawrence Berkeley National Laboratory, 
the National Energy Research Scientific Computing
Center, the University of Oxford, and the Weizmann Institute
of Science;  and the Intermediate Palomar Transient Factory
project, a scientific collaboration among the California Institute of Technology, 
Los Alamos National Laboratory, the University 
of Wisconsin, Milwaukee, the Oskar Klein Center, the
Weizmann Institute of Science, the TANGO Program of the
University System of Taiwan, and the Kavli Institute for the
Physics and Mathematics of the Universe.

The Intermediate Palomar Transient Factory project is a scientific collaboration among the California Institute of Technology, Los Alamos National Laboratory, the University of Wisconsin, Milwaukee, the Oskar Klein Center, the Weizmann Institute of Science, the TANGO Program of the University System of Taiwan, and the Kavli Institute for the Physics and Mathematics of the Universe.  

The PTF database (DR3)  is now publicly available at https:// www.ptf.caltech.edu/news/DR3.  It includes
photometry through Jan. 28, 2015.

Funding for SDSS-III has been provided by the Alfred P. Sloan Foundation, the Participating Institutions, the National Science Foundation, and the U.S. Department of Energy Office of Science. The SDSS-III web site is http://www.sdss3.org/.

SDSS-III is managed by the Astrophysical Research Consortium for the Participating Institutions of the SDSS-III Collaboration including the University of Arizona, the Brazilian Participation Group, Brookhaven National Laboratory, Carnegie Mellon University, University of Florida, the French Participation Group, the German Participation Group, Harvard University, the Instituto de Astrofisica de Canarias, the Michigan State/Notre Dame/JINA Participation Group, Johns Hopkins University, Lawrence Berkeley National Laboratory, Max Planck Institute for Astrophysics, Max Planck Institute for Extraterrestrial Physics, New Mexico State University, New York University, Ohio State University, Pennsylvania State University, University of Portsmouth, Princeton University, the Spanish Participation Group, University of Tokyo, University of Utah, Vanderbilt University, University of Virginia, University of Washington, and Yale University. 

This research has made use of the NASA/IPAC Extragalactic Database (NED) which is operated by the Jet Propulsion Laboratory, California Institute of Technology, under contract with the National Aeronautics and Space Administration.

J.G.C. and B.S. thank  NSF  grant  AST-0908139  to  J.G.C
for  partial  support during the initial early stages of this project. 
S.R.B and K.H. thank the Caltech Summer Undergraduate Research
Fellowship  (SURF) program  for  support.

\clearpage

\begin{deluxetable}{r r r   r r r   r r c}

\tablewidth{0pt}
\tablecaption{Light Curve Parameters for RR Lyr Candidates 
\label{table_lightcurve} }
\tablehead{\colhead{RA} &  \colhead{Dec} & \colhead{Period} & 
       \colhead{0 Phase\tablenotemark{a}} & \colhead{Amp} & \colhead{Mean $R$} &
      \colhead{N($R$)\tablenotemark{b}} & \colhead{N($g$)\tablenotemark{c}} &
         \colhead{Quality\tablenotemark{d} } \\

\colhead{(Deg.)} & \colhead{(Deg.)} & \colhead{(days)} & \colhead{(days)}  
       &  \colhead{($R$ mag)} & \colhead{(mag)} & \colhead{} &
       \colhead{} & \colhead{ } 
 }

\startdata
  3.77632 &   28.37604 &  0.7038231 & 56917.70 &   0.62 &   20.64 &  39 &   3 &   2  \\ 
  10.51398 &   15.64457 &  0.6035999 & 55473.75 &   0.70 &   19.94 &  44 & 131 &   1  \\ 
  13.31085 &   17.13101 &  0.6009332 & 55477.85 &   0.66 &   19.64 &  59 &   0 &   1  \\ 
  21.20049 &   20.43072 &  0.5733125 & 56239.63 &   0.66 &   20.31 &  82 &  16 &   1  \\ 
  21.39914 &    3.82265 &  0.6442016 & 55477.74 &   0.83 &   19.37 & 483 &  78 &   1  \\ 
  22.09602 &   13.81008 &  0.5529058 & 55906.68 &   0.58 &   19.12 & 158 &   0 &   1  \\ 
  26.33246 &   29.49091 &  0.5849030 & 56178.70 &   0.74 &   19.67 & 723 & 178 &   1  \\ 
  28.45696 &   20.34656 &  0.5965452 & 55430.84  &   0.85 &   19.10 & 500 &  91 &   1  \\ 
  29.46073 &   22.95502 &  0.7648931 & 55067.03 &   0.57 &   19.10 &  85 &  11 &   1  \\ 
  29.92910 &   26.09113 &  0.6033530 & 55506.60 &   0.58 &   19.11 &  50 &   0 &   2  \\ 
  32.71250 &   30.72719 &  0.5593969 & 55889.78 &   0.88 &   19.64 &  35 &  51 &   1  \\ 
  82.26892 &    3.38001 &  0.6566145 & 56974.84 &   0.47 &   19.31 &  48 &   0 &   1  \\ 
 115.18660 &   20.78280 &  0.4653525 & 56306.62 &   0.90 &   19.60 & 221 &  37 &   1  \\ 
 120.40829 &   11.03271 &  0.8227144 & 56315.70 &   0.49 &   19.87 &  36 &   0 &   2  \\ 
 125.76719 &   20.88205 &  0.6509836 & 55561.96 &   0.61 &   20.24 & 438 &  91 &   1  \\ 
 133.76732 &   63.42198 &  0.5425147 & 56225.94 &   0.86 &   19.47 &  69 &  33 &   1  \\ 
 135.73433 &   61.63280 &  0.5863690 & 56238.90 &   0.67 &   19.57 & 173 & 100 &   1  \\ 
 143.29643 &   13.29081 &  0.5914273 & 55959.91 &   0.70 &   19.46 & 375 & 138 &   1  \\ 
 153.38060 &   37.91186 &  0.6296058 & 56354.66 &   0.69 &   19.45 &  55 &   0 &   1  \\ 
 158.10118 &    6.32611 &  0.5824709 & 55953.99 &   0.57 &   19.44 &  35 &   0 &   2  \\ 
 163.17050 &   37.51453 &  0.5940006 & 55250.78 &   0.70 &   19.55 & 185 &  77 &   1  \\ 
 164.48911 &   34.74506 &  0.5930000 & 57481.68 &   0.70 &   20.22 &  48 &  45 &   1  \\ 
 180.44336 &   29.98856 &  0.6285709 & 56328.76 &   0.64 &   19.97 & 140 &   4 &   1  \\ 
 180.89786 &   32.49513 &  0.5869673 & 55297.81 &   0.69 &   20.36 & 174 &  86 &   1  \\ 
 181.01610 &   $-$2.64746 &  0.6868450 & 55615.01 &   1.02 &   20.08 &  65 &  87 &   1  \\ 
 185.11430 &   35.10371 &  0.5590766 & 56033.76 &   0.84 &   19.55 &  93 &   0 &   1  \\ 
 185.74637 &   45.19566 &  0.5389892 & 57125.90 &   0.94 &   19.98 & 128 & 131 &   1  \\ 
 188.21655 &    3.81825 &  0.6903981 & 55603.82 &   0.64 &   19.53 & 252 &  46 &   1  \\ 
 190.73627 &   39.43732 &  0.6328233 & 55984.06 &   0.85 &   20.18 &  83 &   8 &   1  \\ 
 191.65421 &   31.93755 &  0.5496644 & 55300.94 &   0.86 &   20.47 & 224 & 100 &   1  \\ 
 193.90346 &   45.04369 &  0.5817848 & 55301.85 &   0.76 &   19.87 &  84 &  97 &   1  \\ 
 196.89822 &   27.27304 &  0.4651977 & 55663.69 &   0.88 &   20.02 & 596 & 118 &   1  \\ 
 197.85255 &   45.04933 &  0.6086488 & 56017.80 &   0.64 &   20.03 & 110 &  94 &   1  \\ 
 198.01382 &   37.50263 &  0.6876892 & 55020.68 &   0.65 &   19.67 & 189 &  33 &   1  \\ 
 198.82190 &   43.19382 &  0.5113264 & 55320.71 &   0.88 &   20.47 & 491 & 129 &   1  \\ 
 199.47290 &   32.11800 &  0.5893799 & 56342.04 &   0.66 &   19.30 & 468 & 224 &   1  \\ 
 201.59912 &   20.38593 &  0.5929822 & 56001.94 &   0.95 &   20.05 &  29 &  17 &   1  \\ 
 203.98247 &   49.90129 &  0.5348190 & 55296.00 &   0.89 &   19.75 &  86 &   0 &   1  \\ 
 204.35236 &   38.22820 &  0.5458603 & 55319.85 &   0.76 &   20.30 & 205 &  43 &   1  \\ 
 205.20570 &   36.85392 &  0.5540681 & 55275.94 &   0.74 &   20.31 & 102 &   1 &   1  \\ 
 205.90588 &   32.55605 &  0.6121942 & 56016.83 &   0.49 &   19.82 & 297 &  41 &   1  \\ 
 206.48932 &   31.08889 &  0.5676502 & 55369.63 &   0.96 &   20.30 &  48 &  19 &   2  \\ 
 207.65234 &   44.81257 &  0.6749544 & 56017.71 &   0.84 &   20.07 & 104 &   5 &   1  \\ 
 209.52818 &   37.12393 &  0.7011946 & 56330.01 &   0.49 &   20.47 & 285 &   5 &   1  \\ 
 210.14015 &   61.58023 &  0.5085552 & 55352.88 &   0.68 &   19.50 & 118 &   0 &   1  \\ 
 210.61269 &   39.29613 &  0.4733887 & 56060.90 &   1.00 &   20.64 & 261 &  53 &   1  \\ 
 210.63774 &   38.23235 &  0.6085635 & 55637.83 &   0.76 &   20.26 & 240 &   9 &   1  \\ 
 211.99417 &   36.80729 &  0.6083931 & 56035.89 &   0.61 &   20.42 & 103 &   8 &   2  \\ 
 212.58949 &   22.25360 &  0.6129632 & 55668.81 &   0.47 &   19.51 &  49 &  82 &   1  \\ 
 215.60762 &   35.91516 &  0.5540740 & 55275.94 &   0.63 &   20.18 & 120 &  13 &   1  \\ 
 216.20433 &   47.13657 &  0.5778699 & 56065.96 &   0.62 &   20.29 &  84 &  71 &   2  \\ 
 218.04994 &   40.74793 &  0.7578443 & 55279.73 &   0.76 &   20.18 & 192 &  54 &   1  \\ 
 218.17694 &   42.63344 &  0.6518604 & 55337.82 &   0.91 &   20.25 & 129 &  18 &   1  \\ 
 223.34651 &    4.97518 &  0.5878099 & 56133.73 &   0.68 &   19.53 &  90 &   5 &   1  \\ 
 223.62280 &   35.96528 &  0.5530584 & 56035.72 &   0.64 &   19.78 & 210 &  25 &   1  \\ 
 226.80437 &   25.50584 &  0.6792830 & 55360.64 &   0.60 &   20.62 &  64 &  13 &   3  \\ 
 229.78807 &   48.06219 &  0.5459260 & 55345.80 &   1.06 &   20.02 &  59 &   0 &   1  \\ 
 230.30850 &   36.28413 &  0.5731339 & 56058.84 &   0.82 &   20.05 &  85 &  11 &   1  \\ 
 231.29987 &   37.24485 &  0.6033793 & 56075.86 &   0.47 &   19.67 & 118 &   0 &   1  \\ 
 233.92378 &   36.95277 &  0.5269989 & 56039.89 &   0.83 &   20.40 &  97 &   8 &   1  \\ 
 236.47386 &   58.07009 &  0.6181521 & 55036.85 &   0.90 &   20.18 & 523 & 180 &   1  \\ 
 239.07004 &   36.43287 &  0.5929263 & 55386.82 &   0.51 &   19.86 & 100 &   5 &   2  \\ 
 239.77188 &   38.56579 &  0.6242449 & 56039.94 &   0.46 &   20.10 & 140 &   0 &   1  \\ 
 240.32909 &   33.09862 &  0.5354092 & 56135.86 &   0.77 &   20.48 & 145 &  15 &   1  \\ 
 242.53758 &   21.51107 &  0.5912649 & 56090.65 &   0.73 &   20.27 &  28 &   1 &   3  \\ 
 242.69720 &   14.62067 &  0.5550773 & 55413.67 &   0.84 &   19.92 & 216 &  28 &   1  \\ 
 246.64397 &    6.24227 &  0.4882272 & 56003.00 &   0.92 &   19.56 &  68 &   0 &   1  \\ 
 246.97688 &   31.55486 &  0.5459153 & 56039.84 &   0.84 &   20.41 & 160 &  56 &   1  \\ 
 247.74931 &   12.76744 &  0.4665125 & 57538.67 &   1.01 &   20.53 & 145 &   0 &   2  \\ 
 248.35423 &   39.41171 &  0.5647064 & 55605.02 &   0.74 &   19.95 & 101 &  29 &   1  \\ 
 251.16412 &   38.47088 &  0.5246175 & 55702.92 &   0.95 &   19.75 & 231 &  73 &   1  \\ 
 253.12921 &   25.36446 &  0.6019734 & 55721.91 &   0.74 &   20.30 &  91 &  26 &   1  \\ 
 256.70535 &   45.84140 &  0.5876117 & 56046.88 &   0.63 &   19.65 & 192 &  36 &   1  \\ 
 257.61386 &   20.88444 &  0.5681136 & 55438.69 &   0.72 &   19.53 & 132 &  30 &   1  \\ 
 258.68228 &   34.30839 &  0.5655935 & 55407.67 &   0.92 &   19.23 & 166 &  52 &   1  \\ 
 258.77704 &   37.91151 &  0.6390487 & 55711.95 &   0.74 &   20.14 & 109 &  35 &   1  \\ 
 313.08780 &    0.13357 &  0.5592707 & 55416.69 &   0.78 &   19.95 & 163 &  66 &   1  \\ 
 318.20523 &    3.33563 &  0.6259214 & 55437.64 &   0.57 &   19.76 &  42 &   4 &   3  \\ 
 320.21838 &    6.36956 &  0.6970433 & 55467.77 &   0.55 &   19.11 &  59 &  12 &   2  \\ 
 321.29257 &    4.89429 &  0.5652243 & 55456.70 &   0.94 &   19.46 &  51 &   0 &   2  \\ 
 323.07294 &   $-$3.48319 &  0.5174600 & 55422.74 &   0.97 &   19.39 & 116 &   0 &   1  \\ 
 323.74008 &   $-$3.52009 &  0.5533006 & 55429.77 &   1.08 &   20.12 &  95 &   0 &   1  \\ 
 327.28622 &   $-$1.67528 &  0.5440798 & 55042.88 &   0.83 &   19.66 & 146 &  14 &   1  \\ 
 328.31757 &    1.61321 &  0.6410172 & 55423.82 &   0.87 &   20.17 & 114 &  19 &   1  \\ 
 329.61261 &   15.67164 &  0.5728618 & 55353.93 &   0.92 &   19.31 & 116 &  17 &   1  \\ 
 331.79990 &   15.46090 &  0.6015934 & 55498.66 &   0.79 &   19.98 & 141 &   5 &   1  \\ 
 331.87677 &   13.06594 &  0.6053543 & 55445.69 &   0.56 &   20.14 & 102 &  27 &   1  \\ 
 332.41092 &   18.24430 &  0.5671536 & 55428.81 &   0.61 &   20.31 &  35 &  16 &   3  \\ 
 332.70486 &   $-$5.19797 &  0.5691122 & 55014.77 &   0.98 &   19.72 &  39 &  30 &   1  \\ 
 332.72507 &    9.88418 &  0.5480604 & 55365.95 &   0.81 &   19.40 & 128 &   0 &   1  \\ 
 332.92017 &   20.10988 &  0.5950155 & 55765.97 &   0.77 &   19.70 &  58 &  20 &   1  \\ 
 333.47647 &   18.46705 &  0.6784748 & 55471.69 &   0.78 &   20.24 &  46 &  20 &   1  \\ 
 334.12289 &   17.94743 &  0.5823412 & 55824.91 &   0.58 &   19.92 &  90 &  42 &   1  \\ 
 334.17758 &   23.86207 &  0.7038231 & 55482.60 &   0.63 &   20.29 &  99 &  15 &   1  \\ 
 334.32181 &    3.08049 &  0.6328018 & 55055.91 &   0.57 &   19.67 &  79 &   0 &   1  \\ 
 334.78058 &   22.16861 &  0.5763521 & 55824.82 &   0.54 &   19.69 & 113 &  19 &   1  \\ 
 334.99887 &   20.37570 &  0.5833233 & 55812.69 &   0.85 &   20.47 & 216 &  12 &   1  \\ 
 336.45602 &   14.85005 &  0.5597054 & 55123.72 &   0.74 &   19.71 & 124 &   0 &   1  \\ 
 339.64056 &    8.46784 &  0.5313083 & 55038.78 &   0.88 &   20.36 &  51 &   0 &   1  \\ 
 343.30026 &   32.76220 &  0.5572263 & 55821.73 &   0.85 &   19.73 &  43 &  80 &   1  \\ 
 344.95831 &    4.83823 &  0.7089223 & 55007.93 &   0.43 &   19.14 & 105 &   0 &   1  \\ 
 346.05371 &    8.77550 &  0.5578092 & 55426.81 &   0.75 &   20.02 &  70 &   0 &   2  \\ 
 346.24451 &    7.73412 &  0.6210057 & 55416.91 &   0.74 &   20.24 & 247 &   0 &   1  \\ 
 346.25085 &   $-$5.34930 &  0.4800212 & 56167.85 &   0.92 &   20.04 & 172 &   0 &   1  \\ 
 348.78879 &   13.35353 &  0.5378452 & 55448.79 &   0.96 &   20.05 & 197 &   0 &   1  \\ 
 349.55212 &   11.92246 &  0.5632662 & 56571.76 &   0.98 &   20.00 & 242 &   0 &   1  \\ 
 350.69650 &   33.79972 &  0.5440564 & 55142.61 &   0.71 &   19.35 & 309 & 385 &   1  \\ 
 351.98282 &   20.68219 &  0.5503856 & 55418.70 &   0.80 &   19.49 &  72 &  10 &   1  \\ 
 354.65012 &   25.68250 &  0.5205241 & 56271.65 &   0.98 &   19.10 & 138 &  89 &   1  \\ 
 355.48550 &   18.87469 &  0.6517739 & 55425.84 &   0.66 &   20.53 &  53 &   9 &   2  \\ 
 356.80493 &    3.27760 &  0.7114636 & 55038.82 &   0.54 &   19.14 &  59 &   0 &   1  \\ 
 358.89981 &   34.25303 &  0.5126550 & 56256.68 &   1.01 &   19.27 &  73 & 197 &   1  \\

  \\
Probable & Sgr & Stream  \\
185.831403  & 11.011716 & 0.5416933 &  55899.95 & 0.76  & \nodata & \nodata &
   \nodata & \nodata \\
189.730895 & 7.902692  & 0.4639002 & 55378.71 & 1.17 &  \nodata  & 287 & 69 & 1 \\
191.481527 & 5.967331 & 0.6101711 & 55333.68  &  0.55 & \nodata  & 280 & 116 & 1 \\
208.767863 & 5.213217 & 0.4890867 & 55330.74 & 1.08 &  \nodata  & 304 & 114 & 1 \\

\enddata

\tablenotetext{a}{Epoch of maximum light in heliocentric Julian date $-$ 2400000 days.
This choice, made for ease of computations, requires 7 digits in the period for
accurate phasing at the present epoch.}

\tablenotetext{b}{Number of epochs taken with the PTF--$R$ filter in which the star was
detected as of late 2016.}

\tablenotetext{c}{Number of epochs taken with the PTF--$g$ filter in which the star was
detected as of late 2016.}

\tablenotetext{d}{Observed $R$ light curve resembles that of a RR Lyr: 1 = excellent, 
2 = probable, 3 = uncertain}

\end{deluxetable}

\clearpage

\begin{deluxetable}{r r r r   r r r}
\tablewidth{0pt}
\tablecaption{$v_r$ for RR Lyr Candidates 
\label{table_vr} }
\tablehead{\colhead{RA} &  \colhead{Dec} & \colhead{Distance\tablenotemark{a}} &
       \colhead{No. Spectra} & \colhead{$v_r$\tablenotemark{b}} & \colhead{$\sigma (v_r)$} & 
      \colhead{Date} \\

\colhead{(Deg.)} & \colhead{(Deg.)} & \colhead{(kpc)}  
       &  \colhead{ } & \colhead{(GSR $km~s^{-1}$)} & \colhead{( $km~s^{-1}$)}
 }
\startdata
   3.77632 &   28.37604 &    109.1 &   1 &    212.7 &   21.0 &  9/2015       \\                     
  10.51398 &   15.64457 &     75.6 &   1 &   $-$240.2 &   17.7 &  9/5/2016     \\                   
  13.31085 &   17.13101 &     65.9 &   1 &   $-$131.1 &   18.4 & 10/2015       \\                   
  21.20049 &   20.43073 &     88.3 &   1 &   $-$124.4 &   18.4 & 10/2015       \\                   
  21.39914 &    3.82266 &     59.2 &   1 &     20.4 &   19.7 & 10/2015       \\                     
  22.09602 &   13.81008 &     50.5 &   1 &    $-$21.2 &   17.8 & 10/2015       \\                   
  26.33246 &   29.49091 &     66.0 &   2 &    $-$38.0 &   12.9 &  dup\tablenotemark{c}  \\          
  28.45696 &   20.34657 &     51.2 &   1 &    $-$34.8 &   19.8 & 10/2015       \\                   
  29.46073 &   22.95502 &     54.9 &   1 &    $-$25.9 &   17.7 & 10/2015       \\                   
  29.92910 &   26.09113 &     51.7 &   1 &     23.6 &   17.8 & 10/2015       \\                     
  32.71250 &   30.72719 &     64.4 &   1 &    $-$56.5 &   20.0 & 10/2015       \\                   
  82.26892 &    3.38001 &     57.8 &   1 &    $-$42.7 &   17.1 &  9/2014       \\                   


 115.18660 &   20.78290 &     60.2 &   1 &   $-$124.7 &   19.3 & 10/2015       \\                   
 120.40829 &   11.03272 &     79.7 &   1 &    224.0 &   17.2 & 10/2015       \\                     
 125.76719 &   20.88206 &     88.6 &   1 &    $-$43.8 &   17.4 &  1/2014\tablenotemark{d}  \\       
 133.76732 &   63.42198 &     59.2 &   1 &     47.9 &   19.8 &  4/2016       \\                     
 135.73433 &   61.63280 &     63.3 &   2 &     11.6 &   13.1 &  dup          \\                     
 143.29643 &   13.29081 &     60.2 &   1 &     99.0 &   18.7 &  4/2016       \\                     
 153.38060 &   37.91186 &     61.1 &   1 &    $-$26.9 &   18.7 &  6/2016       \\                   
 158.10118 &    6.32611 &     59.4 &   1 &   $-$102.7 &   17.8 & 4/2016        \\                   
 163.17050 &   37.51453 &     62.8 &   1 &    $-$59.7 &   18.7 &  4/2015       \\                   
 164.48911 &   34.74506 &     85.5 &   1 &    $-$81.8 &   17.9 &  6/2016       \\                   
 180.44336 &   29.98856 &     77.4 &   1 &    134.3 &   18.3 &  5/2014       \\                     
 180.89786 &   32.49513 &     90.9 &   1 &     54.4 &   18.6 &  5/2014       \\                     
 181.01610 &   $-$2.64747 &     83.4 &   1 &   $-$161.8 &   23.4 &  5/2014       \\                 
 185.11430 &   35.10371 &     61.8 &   1 &    $-$29.7 &   19.7 &  4/2016       \\                   
 185.74637 &   45.19566 &     74.8 &   1 &     49.8 &   20.1 &  4/2016       \\                     
 188.21655 &    3.81826 &     64.9 &   1 &    $-$56.1 &   18.2 &  5/2015       \\                   
 190.73627 &   39.43733 &     85.4 &   1 &     26.3 &   19.8 &  5/2014       \\                     
 191.65421 &   31.93755 &     94.2 &   2 &    $-$48.6 &   14.0 &  dup          \\                   
 193.90346 &   45.04369 &     72.3 &   1 &    $-$36.9 &   19.1 &  5/2014       \\                   
 196.89822 &   27.27304 &     72.9 &   1 &    103.6 &   21.8 &  4/2014       \\                     
 197.85255 &   45.04933 &     79.1 &   1 &    $-$88.0 &   18.3 &  5/2014       \\                   
 198.01382 &   37.50263 &     69.4 &   2 &     98.9 &   12.2 &  dup          \\                     
 198.82190 &   43.19382 &     92.1 &   1 &     29.3 &   20.0 &  4/2014       \\                     
 199.47290 &   32.11801 &     55.9 &   1 &     98.4 &   18.4 &  5/2015       \\                     
 201.59912 &   20.38593 &     79.2 &   1 &    $-$33.7 &   20.5 &  5/2014       \\                   
 203.98247 &   49.90129 &     67.1 &   1 &     56.8 &   20.1 &  5/2014       \\                     
 204.35236 &   38.22820 &     86.6 &   2 &   $-$110.9 &   13.5 &  dup          \\                   
 205.20570 &   36.85392 &     87.3 &   2 &     49.2 &   13.4 &  dup          \\                     
 205.90588 &   32.55605 &     71.8 &   2 &    108.4 &   12.2 &  dup          \\                     
 206.48932 &   31.08890 &     87.7 &   1 &    $-$19.4 &   20.6 &  7/2/2016     \\                   
 207.65234 &   44.81257 &     82.7 &   2 &    $-$73.5 &   15.6 &  dup          \\                   
 209.52818 &   37.12393 &    100.7 &   1 &    $-$11.3 &   17.2 &  4/2014       \\                   
 210.14015 &   61.58023 &     58.9 &   1 &    $-$36.4 &   18.6 &  6/2016       \\                   
 210.61269 &   39.29613 &     97.6 &   1 &    $-$38.6 &   23.9 &  4/2014       \\                   
 210.63774 &   38.23235 &     87.9 &   2 &    $-$60.9 &   14.8 &  dup          \\                   
 211.99417 &   36.80729 &     94.5 &   1 &    $-$52.3 &   18.1 &  4/2014       \\                   
 212.58949 &   22.25361 &     62.2 &   1 &    114.1 &   17.1 &  5/2015       \\                     
 215.60762 &   35.91516 &     82.4 &   1 &     56.5 &   18.2 &  4/2014       \\                     
 216.20433 &   47.13658 &     87.9 &   1 &     17.9 &   18.2 &  5/2014       \\                     
 218.04994 &   40.74793 &     89.9 &   1 &    $-$13.2 &   19.1 &  4/2014       \\                   
 218.17694 &   42.63344 &     89.1 &   1 &   $-$106.9 &   22.1 &  4/2014       \\                   
 223.34651 &    4.97518 &     62.2 &   1 &     $-$4.8 &   18.6 &  5/2015       \\                   
 223.62280 &   35.96528 &     68.6 &   2 &     15.5 &   12.9 &  dup          \\                     
 226.80437 &   25.50584 &    106.9 &   1 &     90.3 &   17.6 &  4/2016       \\                     
 229.78807 &   48.06219 &     76.2 &   1 &    $-$48.8 &   21.3 &  6/2016       \\                   
 230.30850 &   36.28413 &     78.2 &   1 &     71.8 &   19.6 &  5/2014       \\                     
 231.29987 &   37.24485 &     66.6 &   1 &     $-$3.9 &   17.0 &  6/2016       \\                   
 233.92378 &   36.95277 &     90.0 &   1 &     33.6 &   19.6 &  4/2016       \\                     
 236.47386 &   58.07009 &     84.9 &   2 &     33.5 &   19.6 &  dup          \\                     
 239.07004 &   36.43287 &     72.4 &   2 &   $-$175.6 &   16.1 &  dup          \\                   
 239.77188 &   38.56579 &     82.3 &   2 &     64.7 &   12.0 &  dup          \\                     
 240.32909 &   33.09862 &     93.9 &   1 &    $-$38.8 &   19.2 &  4/2014       \\                   
                                     

 242.53758 &   21.51107 &     87.4 &   2 &   $-$190.2 &   13.4 &  dup          \\                   
 242.69720 &   14.62067 &     73.0 &   1 &    106.5 &   19.7 &  6/2016       \\                     
 246.64397 &    6.24227 &     59.8 &   1 &     24.3 &   20.3 &  9/5/2016     \\                     
 246.97688 &   31.55486 &     91.4 &   2 &     81.8 &   16.3 &  dup          \\                     
 247.74931 &   12.76744 &     92.3 &   1 &    171.1 &   21.0 & 5/2014        \\                     
 248.35423 &   39.41171 &     74.7 &   1 &     46.9 &   21.5 &  4/2014       \\                     
                

 251.16412 &   38.47088 &     66.5 &   1 &   $-$152.3 &   20.5 &  5/2014       \\                   
 253.12921 &   25.36446 &     89.2 &   2 &     82.9 &   13.4 &  dup          \\                     
 256.70535 &   45.84146 &     65.5 &   1 &     52.8 &   18.2 &  9/2015       \\                     
 257.61386 &   20.88444 &     61.7 &   2 &    $-$46.0 &   13.3 &  dup          \\                   
 258.68228 &   34.30840 &     53.6 &   1 &     88.9 &   20.3 &  9/2015       \\                     
 258.77704 &   37.91151 &     84.2 &   2 &    $-$83.1 &   13.4 &  dup          \\                   
                  

 313.08780 &    0.13357 &     74.5 &   1 &    $-$27.6 &   19.3 &  9/2014       \\                   
 318.20523 &    3.33564 &     70.3 &   1 &     97.6 &   17.7 &  9/2014       \\                     
 320.21838 &    6.36956 &     53.7 &   1 &    $-$34.6 &   17.6 & 10/2015       \\                   
 321.29257 &    4.89429 &     59.6 &   2 &   $-$118.8 &   14.4 &  dup          \\                   
 323.07294 &   $-$3.48319 &     56.1 &   2 &     10.2 &   14.6 &  dup          \\                   
 323.74008 &   $-$3.52009 &     80.2 &   1 &    $-$98.3 &   21.4 & 10/2015       \\                 
 327.28622 &   $-$1.67528 &     64.7 &   1 &     45.3 &   19.6 &  9/5/2016     \\                   
 328.31757 &    1.61321 &     85.7 &   1 &    $-$21.5 &   18.4 &  9/2014       \\                   
 329.61261 &   15.67164 &     55.7 &   2 &   $-$210.9 &   14.4 &  dup          \\                   
 331.79990 &   15.46091 &     76.9 &   1 &     26.6 &   19.4 & 10/2015       \\                     
 331.87677 &   13.06594 &     82.8 &   1 &   $-$106.4 &   17.7 &  9/2014       \\                   
 332.41092 &   18.24431 &     88.1 &   2 &    $-$72.6 &   12.8 &  dup          \\                   
 332.70486 &   $-$5.19797 &     67.3 &   1 &    141.8 &   20.7 &  9/2014       \\                   
 332.72507 &    9.88418 &     57.4 &   1 &   $-$256.2 &   19.5 & 10/2015       \\                   
 332.92017 &   20.10989 &     67.4 &   1 &     94.4 &   19.2 & 10/2015       \\                     
 333.47647 &   18.46705 &     89.7 &   1 &   $-$155.1 &   19.3 &  9/2014       \\                   
 334.12289 &   17.94743 &     74.1 &   1 &     83.2 &   17.8 &  9/5/2016     \\                     
 334.17758 &   23.86207 &     92.7 &   1 &     37.1 &   18.2 &  9/5/2016     \\                     
 334.32181 &    3.08049 &     67.5 &   1 &   $-$162.4 &   17.8 &  9/2014       \\                   
 334.78058 &   22.16861 &     66.5 &   1 &   $-$101.6 &   17.5 &  9/5/2016     \\                   
 334.99887 &   20.37570 &     95.7 &   1 &     60.8 &   19.8 & 10/2015       \\                     
 336.45602 &   14.85005 &     66.6 &   1 &   $-$238.2 &   19.0 &  9/2014       \\                   
 339.64056 &    8.46785 &     88.5 &   1 &     15.5 &   20.0 & 10/2015       \\                     
 343.30026 &   32.76222 &     67.1 &   1 &     57.1 &   19.8 & 10/2015       \\                     
 344.95831 &    4.83823 &     54.7 &   1 &   $-$237.4 &   16.8 & 10/2015       \\                   
 346.05371 &    8.77550 &     76.8 &   1 &     79.8 &   19.1 & 10/2015       \\                     
 346.24451 &    7.73412 &     87.6 &   1 &   $-$150.4 &   19.0 &  9/5/2016     \\                   
 346.25085 &   $-$5.34935 &     74.4 &   1 &    $-$88.8 &   20.3 & 10/2015       \\                 
 348.78879 &   13.35353 &     77.1 &   1 &   $-$178.3 &   20.5 &  6/2016       \\                   
 349.55212 &   11.92246 &     76.2 &   1 &    $-$75.5 &   20.8 &  6/2016       \\                   
 350.69650 &   33.79972 &     56.1 &   1 &    $-$25.8 &   18.8 & 10/2015       \\                   
 351.98282 &   20.68220 &     59.8 &   1 &   $-$135.7 &   19.4 & 10/2015       \\                   
 354.65012 &   25.68251 &     50.0 &   1 &   $-$106.3 &   20.8 & 10/2015       \\                   
 355.48550 &   18.87470 &    101.4 &   1 &    100.8 &   18.4 &  9/5/2016     \\                     
 356.80493 &    3.27760 &     54.6 &   1 &     $-$7.6 &   17.5 &  9/5/2016     \\                   
 358.89981 &   34.25304 &     53.1 &   1 &    $-$55.9 &   20.9 & 10/2015       \\                   
                 
Sgr & Stream ? \\
185.831403 & 11.011716 & 86.9 & 1 & $-$62.5 & 20.5 & 4/2014 \\
189.730895 & 7.90269   & 86.1 & 1 & $-120.2$ & 25.5 & 4/2014 \\
191.481527 & 5.967331  & 79.5 & 1 & 4.1 & 19.2 & 5/2014 \\
208.767863 & 5.213217  & 83.9 & 1 & $-16.4$ & 29.0 & 4/2014 \\

 \enddata
\tablenotetext{a}{Heliocentric distance.}
\tablenotetext{b}{$v_r$ corrected to the systemic velocity, then to
the heliocentric velocity, and then to the Galactic system of rest.}
\tablenotetext{c}{Two Keck/Deimos spectra have been taken.  See Table~\ref{table_vr_dup} 
for details.}
 \tablenotetext{d}{A DBSP spectrum taken with the Hale Telescope at Palomar Observatory.}

\end{deluxetable}

\clearpage

\begin{deluxetable}{l c rr r}
\tablewidth{0pt}
\tablecaption{$v_r$ for RR Lyr Candidates With Two Keck Deimos Spectra 
\label{table_vr_dup} }
\tablehead{ \colhead{RA} & \colhead{Dec} & 
           \colhead{$v_r$\tablenotemark{a}} & 
    \colhead{$\sigma (v_r)$\tablenotemark{b}} & 
     \colhead{Date} \\
   \colhead{ } & \colhead{(2000)} & 
             \colhead{(GSR: $km~s^{-1}$)} & \colhead{($km~s^{-1}$)}
}
\startdata

Same night \\

135.734329 & 61.632797 & $-$1.6 & 18.5 & 4/5/2016 \\
             &                                & 24.7 & 18.5 & 4/5/2016 \\

204.352357 & 38.228202 & $-$111.6 & 19.1 & 4/5/2016 \\
             &                                & $-$110.3 & 19.1 & 4/5/2016 \\

205.905884    &   32.5560532  &  109.1 & 17.2 & 4/30/2014 \\           
             &                            &  107.7 & 17.2 & 4/30/2014 \\


 323.072937  &   $-$3.48319101 & $-$6.0 & 20.6 & 10/14/2015 \\
             &                              & 26.5 & 20.6 & 10/14/2015 \\

 329.612610   &    15.671644 & $-$217.9 & 20.3 & 10/13/2015 \\
            &                                &  $-$203.9 & 20.3 & 10/13/2015 \\

332.410919  &     18.2443104 & $-$67.6 & 18.1 & 10/13/2015 \\
            &                                  & $-$77.5 & 18.1 & 10/13/2015 \\

~~~ \\

Consecutive & Nights \\

  198.013824   &    37.5026283  & 109.1 & 17.2 & 4/30/2014 \\
            &                            & 88.7 & 17.6 & 5/1/2014 \\

205.205704    &   36.8539162 & 47.3 & 19.0 & 4/30/2014 \\ 
           &                                & 51.0 & 19.1 & 5/1/2014 \\

  207.652344    &   44.8125725 &  $-$89.9 & 22.1 & 4/30/2014 \\             
            &                             &  $-$57.1 & 21.0 & 5/1/2014 \\

 239.070038  &     36.4328651 & $-$181.5 & 22.7 & 4/30/2014 \\               
            &                            & $-$169.8 & 19.7 & 5/1/2014 \\

  246.976883   &    31.5548630 & 84.4 & 23.0 & 4/30/2014 \\                 
           &                               & 79.2 & 23.2 & 5/1/2014 \\

~~~ \\

$\sim$1 Month Apart \\

26.332462 &  29.490912 &   $-$43.3 & 18.2 &  9/19/2015 \\
    &                              &  $-$32.8 & 19.0 &  10/11/2015 \\

    &    & \\

 242.537582   &     21.5110741 & $-$183.8 & 18.9 & 6/2016 \\
     &                         & $-$196.6 & 18.9 & 7/2016 \\

258.777034 &  37.911508 & $-94.1$ & 20.0 & 6/2016 \\
           &            &  $-72.1$ & 20.0 & 9/5/2016 \\

~~~ \\

Nights Separated By &  A Year or More \\
 
191.654206 & 31.937550 &  $-$22.8 & 19.8 & 4/30/2016 \\
   &  &  $-$74.3 & 23.0 & 5/29/2014 \\

 210.637741   &    38.2323532 &  $-$55.3 & 20.9 & 4/30/2014 \\
             &                            &  $-$66.6 & 19.1 & 4/5/2016 \\

223.622803   &    35.9652824  & 25.4 & 18.3 & 5/18/2014 \\
             &                            &  5.6 & 18.3 & 6/4/2016 \\

236.473862   &   58.0700874 &  20.7 & 27.7  &  4/30/2014 \\
            &                            &  46.2 & 20.2  & 6/4/2016  \\

239.771881   &    38.5657883 & 48.9 & 17.0 & 5/28/2014 \\
            &                           & 80.5 & 17.0 & 6/4/2016 \\

253.129211   &    25.3644562 &  78.2 & 19.0 & 4/30/2014 \\
           &                               &  87.7 & 19.0 & 6/4/2016 \\

257.613861  &     20.8844376 &  $-$65.0 & 18.8 & 9/2015 \\
            &                            & $-$27.0 & 18.8 & 4/2016 \\

321.292572    &   4.89428520 & $-$140.7 & 20.4 & 10/2015 \\
           &                           & $-$96.9 & 20.4 & 7/2016 \\
\enddata
\tablenotetext{a}{$v_r$ corrected to the systemic velocity, then to
the heliocentric velocity, and then to the Galactic system of rest.}
\tablenotetext{b}{The 1$\sigma$ uncertainty in the GSR radial velocity.}

\end{deluxetable}

\clearpage

\begin{deluxetable}{r r c  l }
\tablewidth{0pt}
\tablecaption{RR Lyr Candidates That Are Not RR Lyr(ab)\tablenotemark{a} 
\label{table_qso} }
\tablehead{ \colhead{RA} & \colhead{Dec} & \colhead{PTF$-$R}  & \colhead{Comments}
}
\startdata
QSO & (not in NED  & as of 7/2016) \\
3.776320   &  28.376036 & 20.6 & broad em. 5650 \AA \\
 22.518406 &  5.325709 &   20.1  & broad em. 6200, 9200 \AA \\
67.040609  &  0.551808  & 19.9 & broad em. 6600 \AA \\  
328.276107 &  10.135809 & 20.0 & broad em. 5450 \AA \\
338.817911 & 8.405663 &  20.3 & broad em. 6850 \AA \\
  \\
Variable & Stars \\

211.637100 & 20.845664 & 19.9 & variable \\

241.766698 & 22.951423 & 19.8 & variable, period 0.2278 days. \\

250.411946 & 39.1151 & 20.2  & variable \\

291.426967 & 38.535437 & 20.3 & variable \\

 341.865780 & 27.500455 & 20.5 & RRc ? period 0.336 days \\

 344.190408 & $-$5.472125 & 19.7 & SX Phe period 0.0402351 days \\
 355.394934 & 13.261791 & 19.8 & RRc period 0.423 days \\

\enddata

\tablenotetext{a}{Objects which are not $ab$ type RR Lyr based on their
Deimos spectra or on the period derived from their PTF light curves.}

\end{deluxetable}

\clearpage

\begin{deluxetable}{c rrrc}
\tablewidth{0pt}
\tablecaption{Characteristics Of The $v_r$ Distribution
\label{table_vr_sigma} }
\tablehead{ \colhead{$r$ Range} &  \colhead{N} & \colhead{$<v_r(sys)>$\tablenotemark{a}} & 
                 \colhead{$\sigma$\tablenotemark{b}}   \\ 
 \colhead{} & \colhead{ } &
   \colhead{($km~s^{-1}$)} & \colhead{($km~s^{-1}$)\tablenotemark{b})} 
}
\startdata
All \\
$50 < r < 70$  & 51   &  $-27.4$  &  97.4 (95.3)  \\
$70 < r < 85$  & 35   &  $-25.0$  &  103.5 (101.5)  \\   
$50 < r < 85$  & 86   &  $-26.2$  & 99.4 (97.4) \\
$85 < r < 106$ & 26   &  6.2    &  91.9 (89.7)  \\
 ~ \\
$-$ Outliers\tablenotemark{c} \\
$50 < r < 70$ &   47   & $-9.3$ &   78.7 (76.1) \\
$70 < r < 85$ &   33   & $-26.0$  &  89.4 (87.1) \\ 
$50 < r < 85$ &   80  & $-16.2$ & 83.1 (80.7) \\
$85 < r < 109$ &  23  &  $-1.35$ &   68.0 (65.0) \\
\enddata
\tablenotetext{a}{All velocities are in the Galactic standard of rest.}
\tablenotetext{b}{The $\sigma$ values are followed by values in parentheses which have a
   measurement uncertainty of 20~$km~s^{-1}$  removed in quadrature.}
\tablenotetext{c}{Outliers are defined as: ${|}v_r{|} > 170~km~s^{-1}$ ~ for $r > 85$~kpc,
              $> 200~km~s^{-1}$ ~ for $50 < r < 85$~kpc.} 
\end{deluxetable}

\clearpage

\begin{deluxetable}{l rrr}

\tablewidth{0pt}
\tablecaption{Fraction of Candidate RR Lyr That Have Been Confirmed
\label{table_frac_vr} }
\tablehead{ \colhead{$r$ Range} &  \colhead{N(RR Lyr)} & \colhead{N(RR Lyr) $Pr>0.8$ } & 
                 \colhead{Fraction\tablenotemark{a}}   \\ 
 \colhead{(kpc)} &  \colhead{Confirmed RR Lyr} & 
                    \colhead{PTF Candidates} &
     \colhead{(Confirmed/Candidates)}  \\
  \colhead{ } &  \colhead{with $v_r$\tablenotemark{b} } 

}
\startdata
$50 < r < 64$  & 33   & 155  & 0.21  \\
$64 < r < 78$  & 38   &  86  & 0.44 \\
$78 < r < 85$  & 15   & 30   & 0.50 \\
$85 < r < 92$  & 14   & 25   & 0.56 \\
$92 < r < 99$  &  9   & 16   & 0.56 \\
$r > 99$       &  3   &  8   & 0.38 \\
\enddata
\tablenotetext{a}{The ratio of the number of candidate RR Lyr selected from the
PTF database in 2014 with $Pr > 0.8$ to those from this sample with Keck/Deimos $v_r$ for each distance bin.} 
\tablenotetext{b}{The number of RR Lyr in this distance bin from the PTF sample that
have been confirmed with Keck/Deimos spectroscopy and have a $v_r$ given in Table~\ref{table_vr}.}

\end{deluxetable}

\clearpage


\clearpage

\clearpage

\clearpage

\clearpage

%

\clearpage

\clearpage

\clearpage

\clearpage

\clearpage

\clearpage

\clearpage

\clearpage

\clearpage

\clearpage

\end{document}